\documentclass[12pt]{article}

\usepackage{graphicx}
\usepackage{latexsym}
\usepackage{epsf}
\usepackage{epsfig}
\usepackage{amssymb,amsmath}
\usepackage[english]{babel}
\usepackage{amsfonts}
\usepackage{amssymb}
\usepackage{graphics}
\usepackage{mathrsfs}
\usepackage{verbatim}
\usepackage{float}
\def\beq{\begin{equation}}
\def\eeq{\end{equation}}
\def\bea{\begin{eqnarray}}
\def\eea{\end{eqnarray}}
\newcommand{\bal}{\begin{align}}
\newcommand{\eal}{\end{align}}
\def\ba{\begin{array}}
\def\ea{\end{array}}
\def\bi{\begin{itemize}}
\def\ei{\end{itemize}}
\def\ben{\begin{enumerate}}
\def\een{\end{enumerate}}
\def\beq{\begin{equation}}
\def\eeq{\end{equation}}
\def\bc{\begin{center}}
\def\ec{\end{center}}
 \def\bt{\begin{table}}
\def\et{\end{table}}
 \def\btb{\begin{tabular}}
\def\etb{\end{tabular}}

\def\mass2{mass${}^2$}

\def\simlt{\stackrel{<}{{}_\sim}}
\def\simgt{\stackrel{>}{{}_\sim}}

\title{
\begin{flushright}
\normalsize{
ANL-HEP-PR-08-80\\
EFI-08-31\\
FERMILAB-PUB-08-562-T\\}
\end{flushright}
\vspace*{5mm} \Large\textbf{Gauge-Higgs Unification, Neutrino Masses and \\
Dark Matter in Warped Extra Dimensions} \vspace*{1.0cm}
\author{\textbf{Marcela Carena$^{a,b}$, Anibal D.~Medina~$^d$}\\
\textbf{ Nausheen
R.~Shah~$^b$ and Carlos E.M.~Wagner~$^{b,c,e}$}\\
[0.5cm] \normalsize\emph{Theoretical Phys. Dept., Fermi National
Laboratory,
Batavia, IL 60510, USA~$^a$}\\
\normalsize\emph{Department of Physics,
Enrico
Fermi Institute~$^b$}\\
\normalsize\emph{and Kavli Institute
for Cosmological Physics~$^c$,}\\
\normalsize\emph{University of Chicago, 5640 S. Ellis Ave.,
Chicago, IL 60637, USA} \\
\normalsize\emph{Department of Physics, UC Davis, One Shields Ave, Davis, CA 95616~$^d$} \\
\normalsize\emph{HEP Division, Argonne National Laboratory, 9700
Cass Ave., Argonne, IL 60439, USA~$^e$} \\
}}
\date{\today}

\begin{document}
\maketitle
\begin{abstract}
Gauge Higgs Unification in Warped Extra Dimensions provides an
attractive solution to the hierarchy problem. The extension of the
Standard Model gauge symmetry to $SO(5) \times U(1)_X$ allows the
incorporation of the custodial symmetry $SU(2)_R$ plus a Higgs boson
doublet with the right quantum numbers under the gauge group. In the
minimal model, the Higgs mass is in the range 110--150~GeV, while a
light Kaluza Klein (KK) excitation of the top quark appears in the
spectrum, providing agreement with precision electroweak
measurements and a possible test of the model at a high luminosity
LHC. The extension of the model to the lepton sector has several
interesting features. We discuss the conditions necessary to obtain
realistic charged lepton and neutrino masses. After the addition of
an exchange symmetry in the bulk, we show that the odd neutrino KK
modes provide a realistic dark matter candidate, with a mass of the
order of 1~TeV, which will be probed by direct dark matter detection
experiments in the near future.
\end{abstract}
\vspace*{0.5cm} \maketitle

\section{Introduction}

Warped Extra Dimensions present an elegant solution to the hierarchy
problem, where all fundamental parameters are of the order of the
Planck scale. The weak scale--Planck scale hierarchy is obtained by
an exponential warp factor, which is naturally small provided the
Higgs field is localized towards the so-called infrared
brane~\cite{Randall:1999ee}. If all Standard Model (SM) fields
propagate in the bulk, the theory leads to the presence of Kaluza
Klein (KK) modes which tend to be localized towards the IR brane and
therefore couple sizably to the Higgs. This in turn leads to large
mixing between the heavy SM particles and their KK modes, leading to
modifications of the electroweak parameters and therefore to strong
constraints from electroweak precision
measurements~\cite{Davoudiasl:1999tf}--\cite{Hewett:2002fe}. These
constraints may be weakened by the introduction of brane kinetic
terms~\cite{Davoudiasl:2002ua}--\cite{Carena:2004zn} or custodial
symmetries~\cite{Agashe:2003zs},~\cite{Agashe:2006at}, which allow
the presence of KK modes with masses of the order of a few TeV.

One of the attractive features of these models is the natural
explanation of the hierarchy of fermion masses by the localization
of the fermion fields in the
bulk~\cite{ArkaniHamed:1999dc}--\cite{Huber:2000ie}. The chiral
properties of the fermions are obtained by imposing an orbifold
symmetry and demanding that the fields are odd or even under such a
symmetry. Fermion fields that are even under the orbifold symmetry
at the infrared and ultraviolet branes present zero modes, which are
chiral and therefore may be identified with the SM fermion fields.
The localization of the zero modes is governed by the bulk mass
parameter $c k$, with $k$ the curvature of the extra dimension and
$c$, a number of order one. While the zero modes of chiral fields
with $c > 1/2$ couple weakly to the Higgs and are therefore light,
the heavy SM fields are associated with bulk mass parameters $c \leq
1/2$. Due to the exponential behavior of the zero mode wave
functions, large hierarchies between the fermion masses are
generated by small variations of the corresponding $c$-parameters.

Gauge Higgs Unification models identify the Higgs field with the
five dimensional component of the gauge
fields~\cite{gauge:Higgs:unification}. An extended gauge symmetry is
necessary for the successful implementation of this mechanism. In
particular, models based on the gauge group $SO(5) \times U(1)_X$
include the custodial and weak gauge symmetry via $SO(5) \supset
SO(4) \equiv SU(2)_L \times
SU(2)_R$~\cite{Contino:2003ve},\cite{Agashe:2005dk}
\cite{Contino:2006qr}. Moreover, provided the $SO(5)$ symmetry is
broken to a subgroup of $SO(4)$ by boundary conditions at both
branes, the fifth dimensional components of $SO(5)/SO(4)$ have the
proper quantum numbers to be identified with the Higgs field, which
is exponentially localized towards the IR brane.

Since the Higgs originates from gauge fields, its tree level
potential vanishes. In a previous work~\cite{Medina:2007hz}, we
computed the one-loop effective potential and demonstrated that
electroweak symmetry breaking, with the proper generation of
third generation quark and gauge boson masses may be obtained for
the same values of the bulk mass parameters that lead to agreement
with precision electroweak data at the one-loop level. Moreover, we
showed that the Higgs mass is in the range 110--150~GeV and that a
light KK mode of the top quark, $T'$, appears in the spectrum, with a mass
small enough so that the KK gluon may decay into it. The presence of
this light KK mode has a strong impact on the phenomenology of the
model~\cite{Carena:2007tn}. For instance, searches for the KK gluon
by its decay into
top-quarks~\cite{Agashe:2006hk},\cite{Lillie:2007yh},
\cite{Lillie:2007ve} is rendered difficult due to the presence of
the additional dominant decay mode into KK top-quarks and the
broadness of the KK gluon. On the contrary, the constructive
interference between the QCD and KK gluon induced pair production of the
top-quark KK mode allows to search for a $T'$ for values of the masses
much larger than those at reach in the case of just the QCD production
cross section.

In this article, we will analyze the addition of the lepton sector
in the gauge Higgs unification scenario. To add leptons, we will
proceed in a similar way as for the quark sector. The left-handed
leptons will be added in a fundamental representation of $SO(5)$,
with $Q_X = 0$, while the right-handed neutrino and charged lepton
fields will be added in a fundamental representation and a $\bf{10}$
of $SO(5)$ also with charge $Q_X=0$, respectively.

Due to the gauge origin of the Higgs field, a possible local
infrared brane operator $(LH)\;LH/M$, which could lead to large
values of the neutrino Majorana masses, should come from the fifth
dimensional component of the covariant derivative of the lepton
fields and therefore can only naturally arise from the integration
of the right-handed neutrinos, with a local IR Majorana mass.
Indeed, since the fields associated with the right-handed neutrino
zero modes are singlets under the conserved gauge groups on both the
infrared and ultraviolet branes, one can always add Majorana masses
for these fields on the infrared and ultraviolet branes. We will
therefore consider these masses and implement a See-Saw mechanism
for the generation of the light neutrino
masses~\cite{Huber:2002gp},\cite{Huber:2003sf}. We will show how to
incorporate these masses within the context of these models, obtain
the modified profile functions and define the conditions to derive a
realistic spectrum.

Furthermore, the introduction of  an exchange
symmetry~\cite{Panico:2008bx}, which is preserved in the bulk,
yields a natural dark matter candidate in the spectrum, that may be
identified with the odd KK modes of the right-handed neutrino. This
exchange symmetry requires the identification of the bulk mass
parameter of the dark matter candidate with the one of the
right-handed neutrinos, establishing a connection between neutrino
masses and the relic dark matter density. We will show that, if the
odd fermions are assumed to be Dirac particles, the predicted relic
density is the correct one for Dirac fermion masses of the order of
1 TeV. In the Majorana case, somewhat smaller masses are allowed,
and the results depend on the relative values of the ultraviolet and
infrared Majorana masses. We will show that in both the Dirac and
Majorana cases, direct detection experiments will efficiently probe
the existence of the proposed dark matter candidates.

The article is organized as follow: In section 2 we discuss the
properties of leptons in scenarios of Gauge Higgs Unification. In
section 3 we discuss the generation of charged lepton and neutrino
masses. In section 4 we analyze the possibility of incorporating
dark matter via an exchange bulk symmetry, analyzing the couplings
and the annihilation diagrams, as well as the direct detection of
these dark matter candidates. We reserve section 5 for our
conclusions. Some technical details are given in the Appendix.

\section{Leptons in Gauge Higgs Unification Scenarios}

The goal of the current work is to add the lepton sector into the
minimal Gauge Higgs Unification model described above, including
right handed neutrinos, with brane Majorana masses. We will proceed
in a similar way as in the quark case~\cite{Medina:2007hz}, and
discuss the general question of charged lepton and neutrino mass
generation.

Similar to the quark sector, we let the SM $SU(2)_{L}$ lepton
doublet containing the left-handed charged leptons and neutrinos,
$l_L$ and $\nu_L$, arise from a ${\bf 5}_{0}$ of $SO(5)\times
U(1)_X$, where the subscript refers to the $U(1)_X$ charge. The
right handed neutrino, $N_R$ will be included as the singlet
component in a fundamental representation of $SO(5)$, while the
right-handed charged lepton, $l_R$, is placed in a ${\bf 10}_0$ of
$SO(5)$ analogously to the $d_R$. One may also include brane mass
terms connecting different multiplet components, as well as new
brane Majorana masses for the right handed neutrino $N_R$:
\begin{equation}\label{Maj mass}
\Big(M_{UV}\delta(y) - M_{IR}\delta(y-L)\Big)N_R N_R.
\end{equation}

The right handed neutrino, $N_R$, in principle could be identified
with the singlet right-handed component in the same multiplet as the
left-handed leptons. However, in order to naturally suppress lepton
flavor violation effects and maintain agreement with precision
electroweak measurements, the left-handed leptons should be
localized towards the UV
brane~\cite{Carena:2006bn},\cite{Carena:2007ua}. The zero modes of
the corresponding right-handed multiplet will therefore be localized
towards the IR brane. This implies that even with a natural scale
for the brane masses $\mathcal{O}(M_{Pl})$, the exponential
suppression of the wave function at the UV brane would lead to a
effective Majorana mass for $N_R$ which is much smaller than the
Planck scale. This, in turn, after the implementation of the See-Saw
mechanism, leads to too large values for the observed neutrino
masses. Therefore, as we will discuss below, it will prove to be
necessary to have the left-handed leptons in a different multiplet
as the right-handed neutrinos.

If $N_R$ belongs to the same multiplet as the left-handed leptons:
\begin{eqnarray}
\label{1_multiplets}
\begin{array}{c}
\begin{array}{ccccccc}
\xi^{i}_{1L} &\sim& L^{i}_{1L} &=& \begin{pmatrix}
\chi^{e_{i}}_{1L}(-,+)_{1} & l^{n_{i}}_L(+,+)_{0} \\
\chi^{n_{i}}_{1L}(-,+)_{0} & l^{e_{i}}_L(+,+)_{-1}
\end{pmatrix} &\oplus& N^{i}_L(-,-)_{0}~, \vspace{3mm}
\\
\end{array}
\end{array}
\vspace{3mm}
\end{eqnarray}
\begin{eqnarray}
\begin{array}{l}
\xi^{i}_{2R} \sim \\
\\
\begin{array}{ccccccccccc}
T^{i}_{1R} = \begin{pmatrix}
\psi^{\prime i}_R(-,+)_{1} \\
N^{\prime i}_R(-,+)_{0} \\
E^{\prime i}_R(-,+)_{-1} \end{pmatrix} \oplus T^{i}_{2R} =
\begin{pmatrix}
\psi^{\prime\prime i}_R(-,+)_{1} \\
N^{\prime\prime i}_R(-,+)_{0} \\
E^{i}_R(+,+)_{-1} \end{pmatrix} \oplus L^{i}_{2R} =
\begin{pmatrix}
\chi^{e_{i}}_{2R}(-,+)_{1} & l^{\prime \prime n_{i}}_R(-,+)_{0} \\
\chi^{n_{i}}_{2R}(-,+)_{0} & l^{\prime \prime e_{i}}_R(-,+)_{-1}
\end{pmatrix},
\end{array}
\end{array}\nonumber
\end{eqnarray}

Alternatively, the right-handed neutrino can be incorporated in a
different multiplet from the left-handed lepton one. The two
multiplet assignments are as follows:

\begin{eqnarray}
\label{multiplets}
\begin{array}{c}
\begin{array}{ccccccc}
\xi^{i}_{1L} &\sim& L^{i}_{1L} &=& \begin{pmatrix}
\chi^{e_{i}}_{1L}(-,+)_{1} & l^{n_{i}}_L(+,+)_{0} \\
\chi^{n_{i}}_{1L}(-,+)_{0} & l^{e_{i}}_L(+,+)_{-1}
\end{pmatrix} &\oplus& N^{\prime i}_L(-,+)_{0}~, \vspace{3mm}
\\
\xi^{i}_{2R} &\sim& L^{i}_{2R} &=& \begin{pmatrix}
\chi^{e_{i}}_{2R}(-,+)_{1} & l^{\prime {n_{i}}}_R(-,+)_{0} \\
\chi^{n_{i}}_{2R}(-,+)_{0} & l^{\prime {e_{i}}}_R(-,+)_{-1}
\end{pmatrix} &\oplus& N^{i}_R(+,+)_{0}~,
\end{array}
\end{array}
\vspace{3mm}
\end{eqnarray}
\begin{eqnarray}
\begin{array}{l}
\xi^{i}_{3R} \sim \\
\\
\begin{array}{ccccccccccc}
T^{i}_{1R} = \begin{pmatrix}
\psi^{\prime i}_R(-,+)_{1} \\
N^{\prime i}_R(-,+)_{0} \\
E^{\prime i}_R(-,+)_{-1} \end{pmatrix} \oplus T^{i}_{2R} =
\begin{pmatrix}
\psi^{\prime\prime i}_R(-,+)_{1} \\
N^{\prime\prime i}_R(-,+)_{0} \\
E^{i}_R(+,+)_{-1} \end{pmatrix} \oplus L^{i}_{3R} =
\begin{pmatrix}
\chi^{e_{i}}_{3R}(-,+)_{1} & l^{\prime \prime n_{i}}_R(-,+)_{0} \\
\chi^{n_{i}}_{3R}(-,+)_{0} & l^{\prime \prime e_{i}}_R(-,+)_{-1}
\end{pmatrix},
\end{array}
\end{array}\nonumber
\end{eqnarray}
where we show the decomposition under $SU(2)_L\times SU(2)_R$, and
explicitly write the $U(1)_{EM}$ charges. The $L^{i}$s are
bidoublets of $SU(2)_L\times SU(2)_R$, with $SU(2)_L$ acting
vertically and $SU(2)_R$ acting horizontally. The $T^{i}_{1}$'s and
$T^{i}_{2}$'s transform as $({\bf 3}, {\bf 1})$ and $({\bf 1}, {\bf
3})$ under $SU(2)_L\times SU(2)_R$, respectively, while $N^{i}$ and
$N^{\prime i}$ are $SU(2)_L\times SU(2)_R$ singlets. The
superscripts, $i=1,2,3$, label the three generations.

We also show the parities on the indicated 4D chirality, where $-$
and $+$ stands for odd and even parity conditions and the first and
second entries in the bracket correspond to the parities in the UV
and IR branes respectively. Let us stress that while odd parity is
equivalent to a Dirichlet boundary condition, the even parity is a
linear combination of Neumann and Dirichlet boundary conditions,
that is determined via the fermion bulk equations of motion as
discussed below. The boundary conditions for the opposite chirality
fermion multiplet can be read off the ones above by a flip in both
chirality and boundary condition, for example $(-,+)_L\rightarrow
(+,-)_R$. In the absence of mixing among multiplets satisfying
different boundary conditions, the SM fermions arise as the
zero-modes of the fields obeying $(+,+)$ boundary conditions. The
remaining boundary conditions are chosen so that $SU(2)_{L} \times
SU(2)_{R}$ is preserved on the IR brane and so that mass mixing
terms, necessary to obtain the SM fermion masses after EW symmetry
breaking, can be written on the IR brane. Consistency of the above
parity assignments with the original orbifold $Z_2$ symmetry at the
IR brane was discussed in Appendix B of Ref.~\cite{Medina:2007hz}.
The three families will behave similarly, and therefore, we will
drop the family indices and concentrate only on one lepton family.
Large mixing angles in the lepton sector can be naturally obtained
while suppressing lepton flavor changing neutral currents if the
left-handed leptons have similar bulk mass parameters,
$c^i_1$~\cite{Perez:2008ee},\cite{Chen:2008qg}, and in the following
we will assume them to be equal. We will return to this issue in
section~\ref{sec:flavor}. The zero modes of the leptons are too
light and too weakly coupled to the Higgs boson to affect the Higgs
potential in any significant way. The lepton KK modes may be coupled
more strongly to the Higgs, but their gauge invariant mass is much
larger than the Higgs induced one and therefore they tend to
contribute only weakly to both the Higgs potential and to precision
electroweak observables.

One can add localized brane mass terms to the Lagrangian in both the
one and the two multiplet cases:
\begin{eqnarray}
\mathcal{L}_{1} &=& -2\delta(x_{5}-L) \Big[ \bar{L}_{1L} M_{L_2}
L_{2R} + \mathrm{h.c.} \Big] -
\big[M_{IR}\delta(x_5-L)-M_{UV}\delta(x_5)\Big]N_R N_R~;\label{localizedmasses_1}\\
\mathcal{L}_{2} &=& -2\delta(x_{5}-L) \Big[ \bar{N}^\prime_L M_{L_1}
N_R + \bar{L}_{1L} M_{L_2} L_{3R} +  \mathrm{h.c.} \Big] -
\big[M_{IR}\delta(x_5-L)-M_{UV}\delta(x_5)\Big]N_R
N_R~.\label{localizedmasses}
\end{eqnarray}

With the introduction of the brane mixing terms, the different
multiplets are now related via the equations of motion. The
fermions, like the gauge bosons, can be expanded in their KK basis:
\begin{equation}
\psi_{L}(x,x_5) = \sum_n f_{L,n}(x_5,h) \psi_{L,n}(x), \quad \quad
\psi_{R}(x,x_5) = \sum_n f_{R,n}(x_5,h) \psi_{R,n}(x).
\end{equation}
Solving the equations of motion in the presence of $h$ becomes
complicated, as the different modes are mixed. However, 5D gauge
symmetry relates these solutions to solutions with $h =
0$~\cite{Hosotani:2005nz}, with $\Omega(x_5,h)$, the gauge
transformations that removes the vev of $h$:
\begin{equation}
\label{om.tr} \Omega(x_5,h) =
\exp\left[-iC_hhT^{4}\int_0^{x_5}dy\,a^{-2}(y)\right] .
\end{equation}
The form of the bulk profile functions at $h = 0$ is given in
Appendix A.

The boundary masses lead to a redefinition of the effective boundary
conditions for the fermion fields at the branes. Let's first analyze
the case of a Dirac boundary mass term on the infrared brane, $M_L$,
involving fields from  different multiplets $\bar{\Psi}^i_{\rm
L}\Psi^j_{\rm R}$, $i,j =1,2,3$ for the different multiplets. Quite
generally, we shall call the profile of the left-handed field and
the right-handed field participating in the mass term $g_{\rm L}$
and $h_{\rm R}$, respectively. The right handed component,
$\Psi^i_R$ with profile function $g_R$ and the left-handed component
$\Psi^j_L$ with profile function $h_L$ have Dirichlet boundary
conditions on the brane, and therefore $g_R(L) = h_L(L) = 0$. The
equations of motion of the fields are affected by the localized
masses, which induce a discontinuity on the odd-parity profile
functions at the infrared brane. Indeed, keeping only the relevant
terms, the integration of the equation of motion leads to
\begin{equation}
\label{delta} \int^{L+\epsilon}_{L-\epsilon}(\partial_5
g_{R})dx_5= \int^{L+\epsilon}_{L-\epsilon}2M_Lh_{R}\; \delta(x_5-L)dx_5.
\end{equation}
Therefore, we obtain:
\begin{equation}
\label{limit} \lim_{\epsilon\rightarrow 0}
g_{R}(L-\epsilon)= - M_L h_{R}(L),
\end{equation}
and, similarly for the $h_L$

\begin{equation}
\label{limit2} \lim_{\epsilon\rightarrow
0}h_{L}(L-\epsilon)= M_{L_2}g_{L}(L).
\end{equation}
Eq.~(\ref{limit}) and ~(\ref{limit2}) can now be reinterpreted as
the new boundary conditions for the profiles at the IR brane.

Analogously, one can analyze the effect of the Majorana boundary
mass, $M_{i}$, where $i=IR$ or $UV$. Let's take the specific case of
the field $N_R$, with a profile function $h_R$. Its chiral partner
will have a profile function $h_L$ having an odd parity profile on
both branes. The equation of motion in the presence of both the
Majorana masses and the Dirac mass term $M_{L_1}$ leads to the
following relationship
\begin{equation}
\label{delta_i} \int^{y+\epsilon}_{y-\epsilon}(\partial_5
h_{L})dx_5=\int^{y+\epsilon}_{y-\epsilon}\Big[(\pm
2M_{i}h_R-h_L)\delta(x_5-y)-2M_{L_1}g_L\delta(x_5-L)\Big]dx_5.
\end{equation}
where the minus and plus signs are associated with the boundary
conditions at $y=L$, and $y = 0$, respectively. The odd parity at
the branes then implies a Dirichlet boundary condition for the
function $h_{L}$, which as before will present a discontinuity at
the brane. For the IR boundary conditions we obtain:
\begin{equation}
\label{limit2_IR} \lim_{\epsilon\rightarrow
0}h_{L}(L-\epsilon)=M_{IR}h_{R}(L)+M_{L_1}g_L(L).
\end{equation}
For the UV boundary condition, instead:
\begin{equation}
\label{limit2_UV} \lim_{\epsilon\rightarrow
0}h_{L}(0+\epsilon)=M_{UV}h_{R}(0).
\end{equation}
Eq.~(\ref{limit2_IR}-\ref{limit2_UV}) can now be reinterpreted as
the new boundary condition for the profiles at the branes. The generalization
of these expressions to the general case is straightforward.

\subsection{Wave Functions in the Presence of UV Majorana Masses}

The wave functions defined in Appendix A, $\tilde{S}_M$ and
$\tilde{S}_{-M}$ are associated with Dirichlet boundary conditions
at the ultraviolet brane for the left-handed and right-handed
fields, respectively. In the presence of ultraviolet Majorana
masses, however, the boundary conditions for the singlet component
of the fundamental multiplets of $SO(5)$ read
\begin{equation}
f^5_{i,L}(0,0)=M_{UV} f^5_{i,R}(0,0)\label{UV_cond}
\end{equation}
where $i=1,2$ refers to the particular multiplet under analysis. In
order to derive the profile function of these fields, let us first
redefine the functions $a^{3/2} f_{L,R} \to f_{L,R}$ where $a =
\exp(-k x_5)$. With this redefinition, these functions satisfy the
naive normalization condition,
\begin{equation}
\int_0^L dx_5 \; f_n f_m = \delta_{m,n} .
\end{equation}
The general solution for $f_L$ is given by:
\begin{equation}\label{fLgeneral}
f_L(x_5,0)=Aa^{(c-1/2)}\tilde{S}_M+B\left(\frac{a}{z}\right)a^{-(c+1/2)}\tilde{S}_{-M}'.
\end{equation}
where $z$ is the associated particle mass. Defining
\begin{equation}
\tilde{f}_{L,R} = a^{-(c-1/2)} f_{L,R},
\end{equation}
$\tilde{f}_{L,R}$ satisfy the simple equation of motion,
\begin{equation}\label{fLfR}
\tilde{f}_{R,L}(x_5,0)=\mp\frac{a}{z}\partial_5\tilde{f}_{L,R}(x_5,0).
\end{equation}
Now, using Eq.~(\ref{fLfR}), one obtains
\begin{eqnarray}\label{fR}
\tilde{f_R}(x_5,0)&=&-\frac{a}{z}\left(A\tilde{S}_M'-B\left(\frac{a}{z}\right)a^{-2c}\left((1-2c)k\tilde{S}_{-M}'-\tilde{S}_{-M}''\right)\right).
\end{eqnarray}
The second derivative functions may be replaced by means of
the equation of motion of the fermion fields, namely
\begin{equation}\label{sdp}
\tilde{S}_{-M}''=k(1-2c)\tilde{S}_{-M}'-\frac{z^2}{a^2}\tilde{S}_{-M}.
\end{equation}
We therefore see that $\tilde{f_R}$ reduces to:
\begin{equation}\label{fR2}
\tilde{f_R}(x_5,0)=Ba^{-2c}\tilde{S}_{-M}-A\frac{a}{z}\tilde{S}_M'.
\end{equation}
Rewriting these in terms of the $\dot{\tilde{S}}$ rather than
$\tilde{S}'$, with $\dot{\tilde{S}}_{\pm M}=\mp \frac{a(x_5)}{z}
\tilde{S}'_{\pm M} $, we obtain
\begin{eqnarray}
\tilde{f_L}(x_5,0)&=&A\tilde{S}_M+Ba^{-2c}\dot{\tilde{S}}_{-M}\label{fLfR2}\\
\tilde{f_R}(x_5,0)&=&Ba^{-2c}\tilde{S}_{-M}+A\dot{\tilde{S}}_M\label{fLfR3}
\end{eqnarray}

To solve for $A$ in terms of $B$, we need to use the UV boundary
condition induced by the UV Majorana mass for $N_R$,
Eq.~(\ref{UV_cond}):
\begin{eqnarray}\label{ABUV}
Ba^{-2c}\dot{\tilde{S}}_{-M}(0)&=&AM_{UV}\dot{\tilde{S}}_M(0)\nonumber\\
A&=&B\frac{a^{-2c}}{M_{UV}}\frac{\dot{\tilde{S}}_{-M}}{\dot{\tilde{S}}_M}|_{x_5=0}\nonumber\\
\mathrm{since }\quad a=1\quad \mathrm{and}\quad\tilde{S}'_{\pm M}(0,z)=z\quad\mathrm{ for }\quad x_5=0\mathrm{:}&&\nonumber\\
A&=&-\frac{B}{M_{UV}}.
\end{eqnarray}

Therefore, with the coefficients $A$ and $B$ as calculated above,
the singlet functions become
\begin{eqnarray}\label{finalUV}
f^5_{1,L}(x_5,0)&=&C_5  (S_{M_1} - M_{UV} \dot{S}_{-M_1})\\
f^5_{1,R}(x_5,0)&=&C_5   (-M_{UV} S_{-M_1} + \dot{S}_{M_1})
\end{eqnarray}
in the case of a  single multiplet containing the left- and
right-handed neutrinos, and
\begin{eqnarray}
f^5_{2,L}(x_5,0)&=&C_{10} (S_{M_2} - M_{UV} \dot{S}_{-M_2})\\
f^5_{2,R}(x_5,0)&=&C_{10} (-M_{UV} S_{-M_2} + \dot{S}_{M_2})
\end{eqnarray}
in the case of two multiplets, where
$S_{\pm M} = a^{(\pm c - 1/2)} \tilde{S}_{\pm M}$ and
$\dot{S}_{\pm M} = a^{(\pm c - 1/2)} \dot{\tilde{S}}_{\pm M}$.
Therefore, the fermion multiplets with $h = 0$ take the form
\begin{eqnarray}
\label{f.exp.bc.1}
\begin{array}{c}
f_{1,L}(x_{5},0)=\left[\begin{array}{c} C_{1}S_{M_{1}}\\
C_{2}S_{M_{1}}\\ C_{3}\dot{S}_{-M_{1}}\\
C_{4}\dot{S}_{-M_{1}}\\
f^5_{1,L}\end{array}\right]\\
\\
\end{array} & &
f_{3,R}(x_{5},0)=\left[\begin{array}{c} C_{11}S_{-M_{3}}\\
C_{12}S_{-M_{3}}\\ C_{13}S_{-M_{3}}\\ C_{14}S_{-M_{3}}\\
C_{15}S_{-M_{3}}\\ C_{16}S_{-M_{3}}\\
C_{17}S_{-M_{3}}\\C_{18}S_{-M_{3}}\\ C_{19}S_{-M_{3}}\\
C_{20}\dot{S}_{M_{3}}\end{array}\right]
\end{eqnarray}
in the case of a single multiplet containing the left-handed and right-handed
neutrinos, and
\begin{eqnarray}
\label{f.exp.bc}
\begin{array}{c}
f_{1,L}(x_{5},0)=\left[\begin{array}{c} C_{1}S_{M_{1}}\\
C_{2}S_{M_{1}}\\ C_{3}\dot{S}_{-M_{1}}\\
C_{4}\dot{S}_{-M_{1}}\\ C_{5}S_{M_{1}}\end{array}\right]\\
\\
\\
f_{2,L}(x_{5},0)=\left[\begin{array}{c} C_{6}\dot{S}_{-M_{2}}\\
C_{7}\dot{S}_{-M_{2}}\\ C_{8}\dot{S}_{-M_{2}}\\
C_{9}\dot{S}_{-M_{2}}\\
f^5_{2,L}\end{array}\right]
\end{array} & &
f_{3,R}(x_{5},0)=\left[\begin{array}{c} C_{11}S_{-M_{3}}\\
C_{12}S_{-M_{3}}\\ C_{13}S_{-M_{3}}\\ C_{14}S_{-M_{3}}\\
C_{15}S_{-M_{3}}\\ C_{16}S_{-M_{3}}\\
C_{17}S_{-M_{3}}\\C_{18}S_{-M_{3}}\\ C_{19}S_{-M_{3}}\\
C_{20}\dot{S}_{M_{3}}\end{array}\right]
\end{eqnarray}
in the two multiplet case, where
the $C_i$ are normalization constants.

\section{Lepton Spectrum}

Applying the boundary conditions at $x_{5}=L$, taking into account
the mass mixing terms from Eqs.~(\ref{localizedmasses_1}) and
(\ref{localizedmasses}) and using the procedure defined in
Eqs.~(\ref{limit})--(\ref{limit2_IR}), we derive the conditions on
the lepton wave functions $f(L,h)$ in the presence of the Higgs
field. In the case of only one multiplet containing both the
left-handed and right-handed neutrinos one gets the following
conditions at the IR brane:
\begin{eqnarray}
\label{f1.IR.bc}
\begin{array}{ccc}
f_{1,R}^{1,...,4} + M_{L_2} f_{3,R}^{1,...,4}=0 &\quad&
f_{1,L}^{5}-M_{IR}f_{1,R}^{5}=0
\\
&&\\
 f_{3,L}^{1,...,4}-M_{L_2}f_{1,L}^{1,...,4}=0 &\quad& f_{3,L}^{5,...,10}=0
\end{array}
\end{eqnarray}

In the two multiplets, instead, one obtains:
\begin{eqnarray}
\label{f.IR.bc}
\begin{array}{ccccc}
f_{1,R}^{1,...,4} + M_{L_2} f_{3,R}^{1,...,4}=0 &\quad&
f_{1,R}^{5}+M_{L_1}f_{2,R}^{5}=0&\quad& f_{2,L}^{1,...,4}=0
\\
&&&&\\
 f_{3,L}^{1,...,4}-M_{L_2}f_{1,L}^{1,...,4}=0 &\quad& f_{2,L}^{5}-M_{L_1}
f_{1,L}^{5}-M_{IR}f_{2,R}^5=0&\quad& f_{3,L}^{5,...,10}=0
\end{array}
\end{eqnarray}
where the superscripts denote the vector components.

This defines a system of linear equations for the normalization
constants $C_i$. Asking that the determinant of the functional
coefficients of this system vanishes in order to get a non-trivial
solution~\cite{Falkowski:2006vi}, one obtains the following
relations:
\begin{eqnarray}
&\dot{\tilde{S}}_{-M_3}^5 =0&\label{exotic11}\\
&&\nonumber\\
&M_{L_2}^2\tilde{S}_{M_1}\tilde{S}_{-M_3}+\dot{\tilde{S}}_{M_1}\dot{\tilde{S}}_{-M_3}=0&\label{exotic12}\\
&&\nonumber\\
&2\tilde{S}_{M_3}\left[M_{L_2}^2\tilde{S}_{-M_3}\dot{\tilde{S}}_{-M_1}+\tilde{S}_{-M_1}\dot{\tilde{S}}_{-M_3}\right]-M_{L_2}^2\dot{\tilde{S}}_{-M_1}\sin\left[\frac{\lambda h}{f_h}\right]^2=0&\label{lepton}\\
&&\nonumber\\
&2\left[-M_{L_2}^4\tilde{S}_{M_1}\tilde{S}_{-M_3}^2\left[\dot{\tilde{S}}_{-M_1}(\tilde{S}_{M_1}-e^{2c_1kL}M_{UV}\dot{\tilde{S}}_{-M_1})+M_{IR}(1-\tilde{S}_{M_1}\tilde{S}_{-M_1}+e^{2c_1kL}
M_{UV}\tilde{S}_{-M_1}\dot{\tilde{S}}_{-M_1})\right]\right. &\nonumber\\
&-M_{L_2}^2\tilde{S}_{-M_3}\left[2\tilde{S}_{M_1}^2\tilde{S}_{-M_1}-\tilde{S}_{M_1}\left[1+M_{IR}\tilde{S}_{-M_1}\dot{\tilde{S}}_{M_1}-e^{2c_1kL}
M_{UV}\tilde{S}_{-M_1}(2M_{IR}\tilde{S}_{-M_1}-\dot{\tilde{S}}_{-M_1})\right]\right.&\nonumber\\
&\left.-M_{IR}\dot{\tilde{S}}_{M_1}^2\dot{\tilde{S}}_{-M_1}-e^{2c_1kL}
M_{UV}(M_{IR}\tilde{S}_{-M_1}+\dot{\tilde{S}}_{M_1}\dot{\tilde{S}}_{-M_1}^2)\right]\dot{\tilde{S}}_{-M_3}&\nonumber\\
&\left.-\tilde{S}_{-M_1}\left[\dot{\tilde{S}}_{M_1}(\tilde{S}_{M_1}-M_{IR}\dot{\tilde{S}}_{M_1})+e^{2c_1kL}
M_{UV}(1-\tilde{S}_{M_1}\tilde{S}_{-M_1}+M_{IR}\tilde{S}_{-M_1}\dot{\tilde{S}}_{M_1})\right]\dot{\tilde{S}}_{-M_3}^2\right]&\nonumber\\
&+(M_{L_2}^2M_{IR}\tilde{S}_{-M_3}+\dot{\tilde{S}}_{-M_3})\left[M_{L_2}^2\tilde{S}_{-M_3}\left[\tilde{S}_{M_1}+e^{2c_1kL}
M_{UV}\dot{\tilde{S}}_{-M_1}(\tilde{S}_{M_1}\tilde{S}_{-M_1}-\dot{\tilde{S}}_{M_1}\dot{\tilde{S}}_{-M_1})\right]\right.&\nonumber\\
&+\left.\left[e^{2c_1kL}
M_{UV}\tilde{S}_{-M_1}+\dot{\tilde{S}}_{M_1}(\tilde{S}_{M_1}\tilde{S}_{-M_1}-\dot{\tilde{S}}_{M_1}\dot{\tilde{S}}_{-M_1})\right]\dot{\tilde{S}}_{-M_3}\right]\sin\left[\frac{\lambda
h}{f_h}\right]^2=0&\label{neutrino1}
\end{eqnarray}
in the case of a singlet neutrino multiplet, and
\begin{eqnarray}
&\dot{\tilde{S}}_{-M_2}^3=0&\label{exotic1}\\
&&\nonumber\\
&\dot{\tilde{S}}_{-M_3}^5=0&\label{exotic2}\\
&&\nonumber\\
&\left[M_{L_2}^2\tilde{S}_{M_1}\tilde{S}_{-M_3}+\dot{\tilde{S}}_{M_1}\dot{\tilde{S}}_{-M_3}\right]^2=0&\label{exotic3}\\
&&\nonumber\\
&2\tilde{S}_{M_3}\left[M_{L_2}^2\tilde{S}_{-M_3}\dot{\tilde{S}}_{-M_1}+\tilde{S}_{-M_1}\dot{\tilde{S}}_{-M_3}\right]-M_{L_2}^2\dot{\tilde{S}}_{-M_1}\sin\left[\frac{\lambda h}{f_h}\right]^2=0&\label{lepton2}\\
&&\nonumber
\end{eqnarray}
\begin{eqnarray}
&2\left[M_{L_2}^2\tilde{S}_{-M_3}\left[(1-\tilde{S}_{M_1}\tilde{S}_{-M_1})(\tilde{S}_{M_2}-e^{2c_2kL}
M_{UV}\dot{\tilde{S}}_{-M_2})\dot{\tilde{S}}_{-M_2}+M_{IR}(1-\tilde{S}_{M_1}\tilde{S}_{-M_1})\right.\right.&\nonumber\\
&\left.(1-\tilde{S}_{M_2}\tilde{S}_{-M_2}+e^{2c_2kL}
M_{UV}\tilde{S}_{-M_2}\dot{\tilde{S}}_{-M_2})+M_{L_1}^2\tilde{S}_{M_1}\dot{\tilde{S}}_{-M_1}(1-\tilde{S}_{M_2}\tilde{S}_{-M_2}+e^{2c_2kL}
M_{UV}\tilde{S}_{-M_2}\dot{\tilde{S}}_{-M_2})\right]&\nonumber\\
&\tilde{S}_{-M_1}\left[M_{L_1}^2\tilde{S}_{M_1}(1-\tilde{S}_{M_2}\tilde{S}_{-M_2}+e^{2c_2kL}
M_{UV}\tilde{S}_{-M_2}\dot{\tilde{S}}_{-M_2})-\dot{\tilde{S}}_{M_1}\left[\dot{\tilde{S}}_{-M_2}(\tilde{S}_{M_2}-e^{2c_2kL}
M_{UV}\dot{\tilde{S}}_{-M_2})\right.\right.&\nonumber\\
&\left.\left.\left.+M_{IR}(1-\tilde{S}_{M_2}\tilde{S}_{-M_2}+e^{2c_2kL}
M_{UV}\tilde{S}_{-M_2}\dot{\tilde{S}}_{-M_2})\right]\right]\dot{\tilde{S}}_{-M_3}\right]&\nonumber\\
&+\left[M_{L_2}^2\tilde{S}_{-M_3}\left[-2M_{L_1}^2\tilde{S}_{M_1}\dot{\tilde{S}}_{-M_1}-\tilde{S}_{M_2}\dot{\tilde{S}}_{-M_2}+e^{2c_2kL}
M_{UV}\dot{\tilde{S}}_{-M_2}^2\right.\right.&\nonumber\\
&\left.+M_{IR}(-3+2\tilde{S}_{M_1}\tilde{S}_{-M_1}+\tilde{S}_{M_2}\tilde{S}_{-M_2}-e^{2c_2kL}
M_{UV}\tilde{S}_{-M_2}\dot{\tilde{S}}_{-M_2})\right]&\nonumber\\
&\left.+\left[2M_{IR}\tilde{S}_{-M_1}\dot{\tilde{S}}_{M_1}-M_{L_1}^2
(1+2\tilde{S}_{M_1}\tilde{S}_{-M_1}-\tilde{S}_{M_2}\tilde{S}_{-M_2}+e^{2c_2kL}
M_{UV}\tilde{S}_{-M_2}\dot{\tilde{S}}_{-M_2})\right]\dot{\tilde{S}}_{-M_3}\right]\sin\left[\frac{\lambda
h}{f_h } \right]^2&\nonumber\\
&+\left[M_{L_2}^2
M_{IR}\tilde{S}_{-M_3}+M_{L_1}^2\dot{\tilde{S}}_{-M_3}\right]\sin\left[\frac{\lambda
h}{f_h } \right]^4=0&\label{neutrino2}
\end{eqnarray}
in the case of two multiplets. In the above, for simplicity, we did
not write the dependence on $L$ and $z$ and furthermore, we have
used the Crowian:
\beq \label{e.cr}
-\dot{\tilde{S}}_M(x_5,z)\dot{\tilde{S}}_{-M}(x_5,z) +
\tilde{S}_M(x_5,z) \tilde{S}_{-M}(x_5,z) = 1 \, . \eeq
The roots of the above equations define the values of $z$
corresponding to the masses of the lepton zero modes and KK modes in
the presence of the Higgs fields.

\subsection{Charged Lepton Masses}

Since the charged lepton masses are given by the mixing of the first
and third multiplet via $M_{L_2}$, the expression determining its
mass is formally the same for the case in which the right-handed
neutrino is in the same multiplet as the left-handed one as for the
two neutrino multiplet case (Eqs.~(\ref{lepton}) and
(\ref{lepton2})). Additionally, since the lepton masses are much
smaller then $\tilde{k}$, one can use an expansion of $\tilde{S}_M$
for small values of $z/\tilde{k}$. As we shall show, the approximate
functions we derive in this way agree very well with the full
numerical investigation we carried out. We shall concentrate on
values of $c_1 \simgt 0.5$, which are preferred to cancel flavor
violation effects in a natural way and provide agreement with
precision electroweak data~\cite{Carena:2007ua},\cite{Perez:2008ee}.

At small values of $z$ compared to $\tilde{k}$, one can express the
function $\tilde{S}_M$ in the form~\cite{Falkowski:2006vi}:
\begin{equation}\label{zapprox}
\tilde{S}_{M}\approx z\int^{x_{5}}_{0}
a^{-1}(y)e^{-2My}dy+\mathcal{O}(z^{3}) \, .
\end{equation}

Using this in Eq.~(\ref{lepton}), we can solve for the mass:

\begin{equation}\label{emass full}
\left(\frac{z}{\tilde{k}}\right)=\frac{M_{L_2}e^{(\frac{1}{2}+c_3)kL}
\sin[\frac{\lambda
h}{f_h}]\sqrt{2 \; (\frac{1}{2}-c_1)(c_3^2-\frac{1}{4})}}{\sqrt{
\left[(\frac{1}{2}-c_3)(e^{2(c_1-\frac{1}{2})kL}-1)-M_{L_2}^2
(c_1-\frac{1}{2})(e^{2(c_3-\frac{1}{2})k}-1)\right]
(e^{2(\frac{1}{2}+c_3)kL}-1)}}.
\end{equation}

\begin{figure}[!t]\centering\scalebox{1}[1]
{\includegraphics[width=0.8\textwidth,bb=1cm 1cm 20cm
18cm]{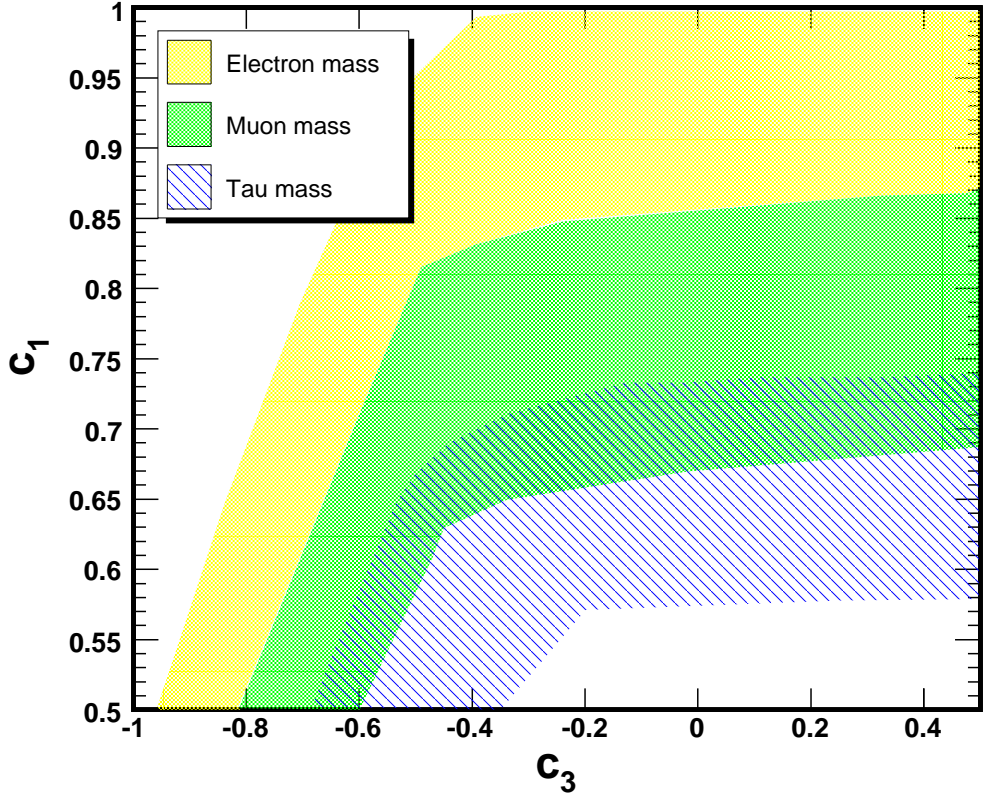}}\caption{\footnotesize{Region of $c_1$, $c_3$
parameter space consistent with the known charged lepton masses.
The bands correspond to variations of the values
of the $\tilde{k}$ and $M_{L_2}$ parameters
in the range 1.5~TeV~$\simlt \tilde{k} \simlt$~5.5~TeV and
$0.1 \simlt M_{L_{2}} \simlt 5$.}}
\label{figure:lepmass}
\end{figure}

If $c_1>0.5$ and $-0.5<c_3<c_1$, this reduces to:

\begin{equation}
\label{c3g0.5lc1}
\left(\frac{z}{\tilde{k}}\right)=M_{L_2}e^{(\frac{1}{2}-c_1)kL}
\sin\left[\frac{\lambda
h}{f_h}\right]\sqrt{2 \; (c_1-\frac{1}{2})(c_3+\frac{1}{2})}
\end{equation}
For $c_1>0.5$ and $c_3>c_1$, instead,
\begin{equation}
\label{c3gc1}
\left(\frac{z}{\tilde{k}}\right)=e^{(\frac{1}{2}-c_3)kL}\sin\left[\frac{\lambda
h}{f_h}\right]\sqrt{2 \; (c_3^2-\frac{1}{4})}
\end{equation}
Finally, for the case $c_1>0.5$ and $c_3<-0.5$:

\begin{equation}
\label{c3l0.5}
\left(\frac{z}{\tilde{k}}\right)=M_{L_2}e^{(1+c_3-c_1)kL}
\sin\left[\frac{\lambda h}{f_h}\right]\sqrt{2
\;(\frac{1}{2}-c_1)(c_3+\frac{1}{2})}
\end{equation}
where we have assumed that $M_{L_2} \neq 0$. We note that in the
above, the lepton masses depend at most linearly on $M_{L_2}$. As
shown in Fig.~\ref{figure:lepmass}, the above relations are
verified by our numerical results.  Realistic lepton masses may be
obtained for e.g. for $c_1 \simeq 0.6$ and $c_3 \simeq -0.55$,
$-0.65$ and $-0.8$ for the case of the tau, muon and electron,
respectively. If a common value of $c_1$ for the three generations
is demanded, as explained above,  and for values of $M_{L_2}$ of
order one, as chosen in Fig.~\ref{figure:lepmass},
the value of $c_1$ is restricted to
be in the range $0.5 \simlt c_1 \simlt 0.75$. Larger values of $c_1$
become incompatible with the heavier charged lepton masses.

\subsection{Neutrino Masses}

The Neutrino masses are analyzed in a similar manner. First we look
at the case in which the left-handed and right-handed neutrinos
belong to the same muliplet case. For $c_1>0.5$,
\begin{equation}\label{1mn2}
\left(\frac{z}{\tilde{k}}\right)=\frac{(c_1-\frac{1}{2})e^{2(\frac{1}{2}-c_1)kL}\sin[\frac{\lambda
h}{f_h}]^2}{M_{IR}}
\end{equation}
From Eq.~(\ref{1mn2}) we see that values of $c_1\sim 1$ would be
necessary in order to get the correct values for the neutrino
masses. However, values of $c_1 \simeq 1$ are strongly constrained
in order to reproduce the proper charged lepton masses. Indeed, as
we emphasized at the end of last section, the proper values of
$\tau$ and $\mu$ masses may not be obtained for $c_1\sim 1$.
Therefore, we conclude that if all $c_1$'s are about the same, as
preferred to obtain large flavor mixing naturally without inducing
large lepton flavor changing effects~\cite{Perez:2008ee}, two
multiplets are required in order to obtain the correct lepton
spectrum.

In the two multiplet case, the dependence of the neutrino masses on
the mixing with the third multiplet through $M_{L_2}$ is always
exponentially suppressed. Therefore, we shall set $M_{L_2}=0$ in the
following approximate expressions. The approximate mass expressions,
for
\begin{itemize}
\item $c_1>0.5$ and $c_2> 1/(k\;L)$:

\begin{equation}\label{2mn1}
\left(\frac{z}{\tilde{k}}\right)=\frac{M_{L_1}^2
(c_1-\frac{1}{2})e^{2(\frac{1}{2}-c_1)kL} \sin\left[\frac{\lambda
h}{f_h}\right]^2}{M_{IR}}
\end{equation}

\item $c_1>0.5$ and $c_2< -1/(k\;L)$:

\begin{equation}\label{2mn2}
\left(\frac{z}{\tilde{k}}\right)=\frac{M_{L_1}^2
(c_1-\frac{1}{2})e^{2(\frac{1}{2}-c_1+c_2)kL}
\sin\left[\frac{\lambda h}{f_h}\right]^2}{M_{UV}}
\end{equation}

\item $c_1>0.5$ and $c_2\sim 0$:

\begin{equation}\label{2mn3}
\left(\frac{z}{\tilde{k}}\right)=\frac{M_{L_1}^2 (c_1-\frac{1}{2})
e^{2(\frac{1}{2}-c_1)kL} \sin\left[\frac{\lambda
h}{f_h}\right]^2}{M_{UV} + M_{IR}}
\end{equation}
\end{itemize}
where in Eq.~(\ref{2mn3}) we have assumed $M_{UV} \neq - M_{IR}$.

\begin{figure}[!t]\centering\scalebox{0.8}[0.8]
{\includegraphics[width=\linewidth,bb=2 25 518 465,
clip=]{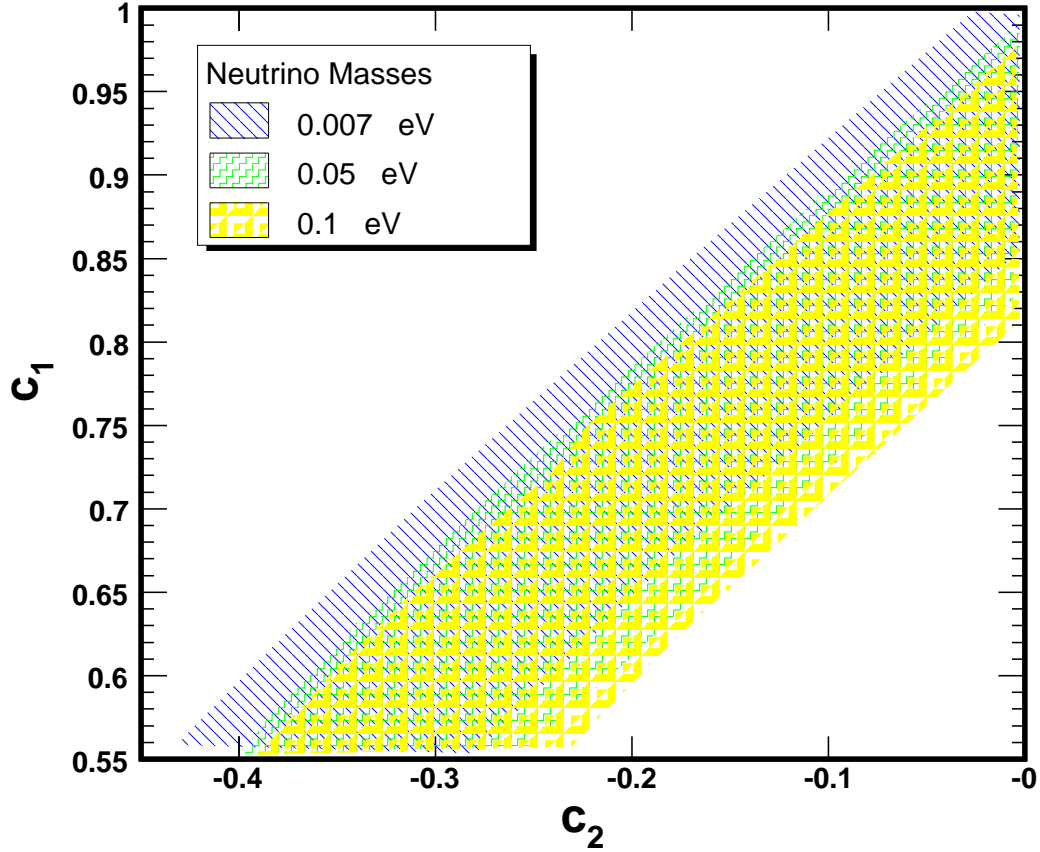}}\caption{\footnotesize{Region of $c_1$, $c_2$
parameter space consistent with the neutrino masses of interest:
$c_1>0.5$ and $-0.5<c_2<0$.
The bands
correspond to variations of the values of the parameters
$\tilde{k}$, $M_{L_1}$ and $M_{UV,IR}$ in the range
1.5~TeV~$\simlt \tilde{k} \simlt$~5~TeV, $0.1 \simlt M_{L_{1}} \simlt 1.5$
and $0.5 \simlt M_{IR,UV} \simlt 2.5$.   }}
\label{fig.n1}
\end{figure}

In the linear regime, $(\lambda h/f_h)^2 \ll 1$, these neutrino
masses become proportional to the square of the Higgs vev, and show
the characteristic See-Saw behavior governed by the brane Majorana
masses. From the above expressions we see that it will only be
possible to generate the correct order of the neutrino masses when
$c_2 \simgt -0.4$. Moreover, for $c_2\gtrsim 0$, the values of $c_1$ are
such that the correct heavy lepton masses cannot be generated. These
conclusions are verified in our numerical work. We present the
relevant parameter space in the $c_1-c_2$ plane leading to the
correct order of the neutrino masses in Fig.~\ref{fig.n1}. The width
of the bands for the different masses corresponds to varying
$\tilde{k}$ and the different brane masses in the range indicated
in Fig.~\ref{fig.n1}.  As indicated by the
above expressions, we were not able to numerically find any
solutions for $c_2<-0.5$. Finally, although positive values of $c_2$
are not represented in Fig.~\ref{fig.n1}, the neutrino masses become
independent of $c_2$ for $c_2>0$, and therefore the values of $c_1$
are the same as for $c_2 = 0$.

\subsection{Flavor Problem}
\label{sec:flavor}

In the above, we have not discussed the problem of flavor. It is
well known that, if the effective Yukawa couplings have an anarchic
structure, large flavor violating effects are induced, which may
only be suppressed by pushing the KK masses to values above 10~TeV,
excluding any possible phenomenology of warped extra dimensional
models at the LHC, as well as any possible dark matter candidate
coming from the KK modes (see, for example Ref.~\cite{Csaki:2008zd}
as well as Ref.~\cite{Casagrande:2008hr} for an alternative approach
to this question). The problem stems in part from the fact that the
Yukawa couplings and the bulk mass parameters are not diagonalized
in the same basis, and therefore the quark mass eigenstates have
flavor violating couplings with the gluon KK modes. A possible
solution to this problem is to demand an alignment between the bulk
mass parameters and the Yukawa couplings, as has been proposed in
Ref.~\cite{Fitzpatrick:2008zza}. Flavor violation in the lepton
sector can also be suppressed by a similar
alignment~\cite{Perez:2008ee},\cite{Chen:2008qg}. This is equivalent
to demanding that the bulk mass parameters obey the following
relationships:
\begin{eqnarray}
c_3 & = & I + a_3 k^2 Y_l^{\dagger} Y_l
\nonumber\\
c_2 & = & I + a_2 k^2 Y_\nu^{\dagger} Y_\nu
\nonumber\\
c_1 & = & I + a_l k^2 Y_l Y_l^{\dagger} + a_\nu k^2 Y_\nu Y_\nu^{\dagger};
\end{eqnarray}
where $Y_l$ and $Y_\nu$ are the effective charged lepton and
neutrino Yukawa couplings and $a_l$, $a_\nu$, $a_2$ and $a_3$ are
numerical constants and the $c_i$ are now matrices where the
off-diagonal terms mix the different generations.

The Gauge-Higgs unification structure introduced above demands a
redefinition of the above equations, since no explicit Yukawa
coupling has been written. As can be seen from
Eqs.~(\ref{c3g0.5lc1})--(\ref{c3l0.5}), and
Eqs.~(\ref{2mn1})--(\ref{2mn3}), the role of the Yukawa coupling is
now being played by the boundary masses $M_{L_1}$ and $M_{L_2}$.
Hence, the above equations must be rewritten as
\begin{eqnarray}
c_3 & = & I + a_{32} M_{L_2}^{\dagger} M_{L_2}
\nonumber\\
c_2 & = & I + a_{21} M_{L_1}^{\dagger} M_{L_1}
\nonumber\\
c_1 & = & I + a_{12} M_{L_2} M_{L_2}^{\dagger}
+ a_{11} M_{L_1} M_{L_1}^{\dagger};
\end{eqnarray}
If $a_{12} \gg a_{11}$, the charged lepton masses would be
diagonalized in the same basis as the bulk mass parameters, inducing
minimal flavor violation in the lepton sector. In this case, all
flavor violation will be associated with the charged currents,
leading to values of the lepton flavor violation processes
consistent with experiment for KK masses as low as a few TeV. As
emphasized above, large mixing angles may naturally arise within
this framework, if all left-handed zero modes localization
parameters take equal values, namely when $a_{12} \simeq a_{11}
\simeq 0$~\cite{Perez:2008ee}.

The results of Refs.~\cite{Perez:2008ee},\cite{Chen:2008qg} show
that it is possible to impose a flavor symmetry in this class of
models such that no large flavor violation occurs. In this work, we
will assume that this, or a
similar~\cite{Csaki:2008qq},\cite{Csaki:2008eh},\cite{Santiago:2008vq}
flavor protection exists, and will postpone a more detailed analysis
of this question for future work.

\section{Dark Matter}

Dark Matter in warped extra dimensions was first introduced in
Ref.~\cite{Agashe:2004ci} within a framework which solves the proton
stability problem in unification scenarios. The introduction of a KK
parity in warped extra dimensions, leading to a stable dark matter
candidate, was further proposed in Ref.~\cite{Agashe:2007jb}. In
this work, we shall proceed in a different way: Following
Ref.~\cite{Panico:2008bx}, we shall introduce an additional exchange
$Z_2$ symmetry under which all the lepton multiplets introduced so
far would be even. One can then define extra fermion multiplets,
that will be chosen as the ``odd'' partners of the lepton
multiplets. If this symmetry is preserved, the lightest odd particle
(LOP) will be stable, and therefore can be considered as a possible
dark matter candidate. In the framework of Ref.~\cite{Panico:2008bx}
the equality of the even and odd mass parameters was enforced by
giving the original fermions, whose even and odd combinations form
the even and odd fields, different charges under an extended
$U(1)_{X_1} \times U(1)_{X_2}$ gauge symmetry. Since in our case the
leptons are neutral under $U(1)_X$ this property does not hold.
Additionally, contrary to Ref.~\cite{Panico:2008bx}, we shall assume
that the quark and gauge boson multiplets do not have odd partners.

Even though, the structure of our model does not require the
equality of the bulk masses for the odd and even fields, for
simplicity, we shall assume that the bulk mass parameters are
identified with each other and are controlled by the requirement of
obtaining the correct small neutrino masses via the see-saw
mechanism. Our assumption is equivalent to requiring that there are
no off-diagonal bulk mass parameters mixing the original fields for
which the $Z_2$ exchange symmetry holds.

In order to explore this possibility, we shall identify the
multiplet containing the dark matter candidate with the odd partners
of the second lepton multiplet containing the right-handed
neutrinos. As has been shown in the previous section, this demands
values of $c_2 \simlt 0$, and therefore we shall require the bulk
mass parameter of the dark matter candidate to be in this range.

The exchange symmetry, introduced in Ref.~\cite{Panico:2008bx}
allows arbitrary boundary masses between even fields, necessary for
obtaining the proper lepton masses, as well as between the odd
fields. Boundary masses mixing odd and even fields are, instead,
forbidden by this symmetry. On the other hand, the boundary
conditions for the even and odd fields are independent of each
other. Therefore, the main link between even and odd fields is
through the identification of the bulk mass parameters. For
simplicity, we shall choose the boundary conditions of the odd
partners of the chiral first and third lepton multiplets, containing
the left-handed and right-handed charged leptons, to be the same as
the one of the even biodoublet components, (- +) and (+ -),
respectively. For values of the bulk masses $c_1 > 0.5$ and $c_3 <
-0.5$, these fields will be relatively heavy, with masses about a
few times~$\tilde{k}$, and decoupled from the Higgs.

For the second multiplet, the odd fields, denoted by $\xi_o$ will
have the same boundary conditions as $\xi_2$ for the even
bidoublets. However, since the fifth component does not contain a
zero mode, it must have different boundary conditions on the IR and
UV branes. That leaves us with two options for the left handed
component of this field: $(+,-)$ or $(-,+)$. The goal here is to
consider the possibility of a neutral odd lepton, mainly singlet
under the $SU(2)_L \times SU(2)_R$ symmetry, as a dark matter
candidate. The singlet and the doublet states mix via their
interactions with the Higgs field, which will act as a small
perturbation to their masses. In order to make the coupling to the
Higgs effective and to split the doublet and singlet masses, we
shall choose the singlet right-handed field to have the same
boundary conditions on the IR brane as the bidoublet left-handed
field for at least one of the three odd partners of the second
multiplets. Therefore, the boundary conditions for the odd multiplet
containing the LOP are chosen to be:
\begin{eqnarray}
\label{oddmult}
\begin{array}{c}
\begin{array}{ccccccc}
\xi^{o}_{R} &\sim& L^{o}_{R} &=& \begin{pmatrix}
C^{o}_{R}(-,+)_{1} & n^{\prime {o}}_R(-,+)_{0} \\
n^{o}_{R}(-,+)_{0} & C^{\prime {o}}_R(-,+)_{-1}
\end{pmatrix} &\oplus& N^{o}_R(+,-)_{0}~,
\end{array}
\end{array}
\vspace{3mm}
\end{eqnarray}
Regarding the other two generations of odd partners of the second
multiplets, for simplicity, we will choose their singlet states to
have the opposite boundary conditions from the one presented in
Eq.~(\ref{oddmult}). This would force their masses to be heavy for
$c_2 \leq 0$, ensuring that the multiplet with the boundary
conditions given by Eq.~(\ref{oddmult}), $\xi_o$, would generate the
LOP. In addition, small Dirac boundary masses may be included, which
would allow a small mixing between the odd multiplets inducing
decays of the heavier generation odd states to the LOP, through the
weak gauge bosons and the Higgs boson. Even in the case of a very
small mixing, due to the large mass differences, the lifetime of
these heavy odd partners would naturally be very short, and
therefore these heavier odd multiplets would not contribute to the
LOP relic density in any relevant way.

Finally, in order to estimate the dark matter density, we shall
restrict ourselves to the first level of odd KK modes, since they
give the dominant contribution to the annihilation cross section.
This is due not only to their relatively small masses with respect
to the heavier modes in the tower, but also due to their larger
couplings to the LOP. We have checked that the inclusion of the
second KK level leads to a very small modification of the
annihilation cross section and therefore of the freeze-out
temperature and the predicted relic density. Let us finally comment
that in the region consistent with the proper dark matter density,
the mixing between the singlet and doublet particles is small and
these particles lead to only a small contribution to the Higgs
effective potential and the precision electroweak observables.

Similar to the standard model fields discussed in the previous
section, the boundary conditions, Eq.~(\ref{oddmult}), lead to a set
of equations which determine the masses of our odd multiplet
fermions. For the KK modes that couple to the Higgs boson, in the
case of vanishing Majorana masses for the odd fields, we find the
following condition
\begin{equation}\label{odddet}
\sin\left[\frac{\lambda
h}{f_h}\right]^2+\dot{\tilde{S}}_{M_2}\dot{\tilde{S}}_{-M_2}=0.
\end{equation}

\begin{figure}[!t]\centering\scalebox{0.65}[0.65]
{\includegraphics[width=\linewidth,bb=11 21 518 467,
clip=]{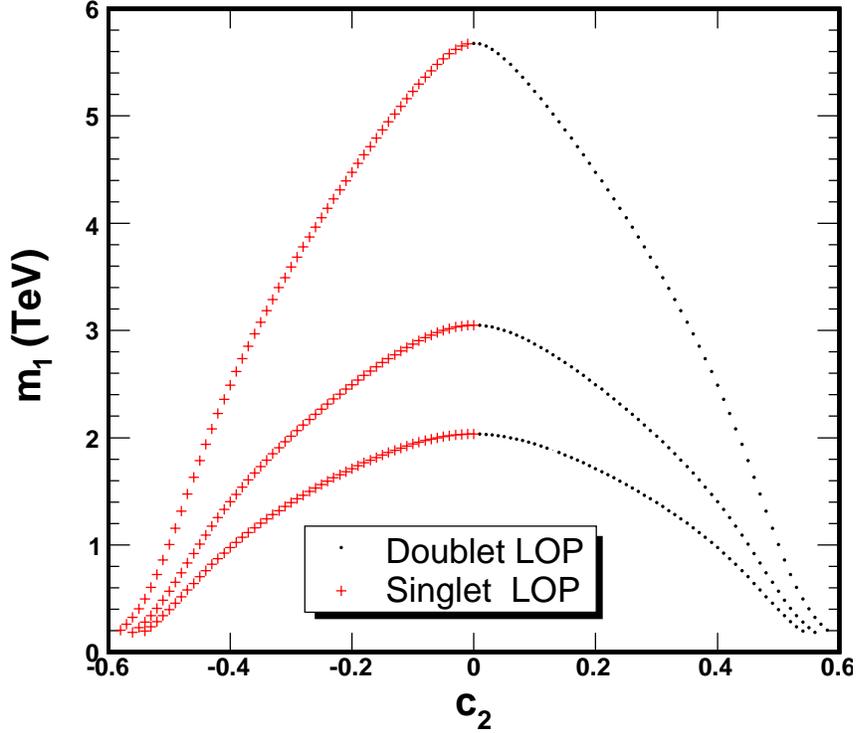}}\caption{\footnotesize{The Dirac mass of the LOP,
$m_1$ as a function of $c_2$, the localization parameter for the odd
fermions for three values of $\tilde{k}=~1.5, ~2.2$ and $3.8$ TeV.
}} \label{fig.cm1D}
\end{figure}

The solutions for the LOP are plotted in Fig.~\ref{fig.cm1D}.
The behavior of the LOP mass may be understood from the $h = 0$
limit. In this limit, the singlet and doublet states don't mix and
the singlet becomes the lightest odd particle for $c_2 < 0$, while
the doublet becomes the LOP for $c_2 > 0$. At $c_2=0$ and $h=0$, the
singlet and doublet masses are degenerate. When $h \neq 0$ the
singlet and doublet states mix and their masses are split by the
Higgs v.e.v. Only the lightest state mass is plotted in
Fig.~\ref{fig.cm1D}. In the presence of the Higgs, at $c_2 = 0$, the
LOP is an equal admixture of the singlet and doublet state. As we
move away from $c_2=0$, the roots of the determinant will start
splitting into two clearly spaced masses.  For $c_2<0$, the lighter
mass is mostly a singlet state and the heavier one is mostly a
doublet state. Since these are Dirac particles, the positive and
negative roots of the determinant are equal.

\subsection{Odd Majorana Masses}

In the above, we have not considered the impact of Majorana mass
terms that could in principle be written for this multiplet, both on
the IR and the UV branes, as was done for the even neutrinos, and
would modify the couplings to the Higgs boson. Including the
Majorana masses for the odd multiplet, the equation determining the
odd lepton masses is given by
\begin{eqnarray}\label{odd_maj}
&\dot{\tilde{S}}_{-M_2} \left(\dot{\tilde{S}}_{M_2}-M_{IR_o}
\tilde{S}_{M_2}-e^{2 c_2 k L} M_{UV_o}
\left(\tilde{S}_{-M_2}-M_{IR_o}
\dot{\tilde{S}}_{-M_2}\right)\right)+\sin\left[\frac{\lambda h}{f_h}\right]^2 =0&\nonumber\\
&&
\end{eqnarray}

\begin{figure}[!t]\centering\scalebox{0.75}[0.75]
{\includegraphics[width=\linewidth,bb=11 21 568 515,
clip=]{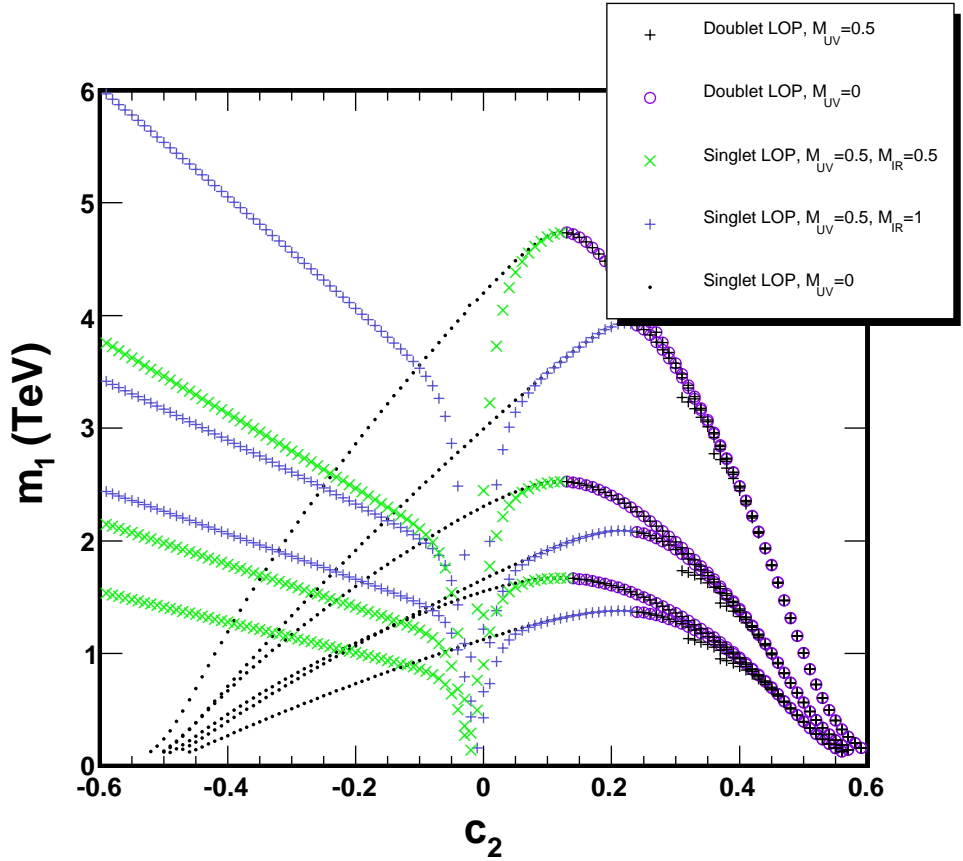}}\caption{\footnotesize{The Majorana mass of the
LOP, $m_1$ as a function of $c_2$, the localization parameter for
the odd fermions, with different values of $M_{UV}$ marked, and for
three values of $\tilde{k}=~1.5, ~2.2$ and $3.8$ TeV (corresponding
to the three convergent purple circle lines from bottom to top for
$c_2>0$), and two different values of $M_{IR_o} = ~1$ and $0.5$ for
each $\tilde{k}$ and $M_{UV_o}$ (from top to bottom for $c_2 < 0$).
}} \label{fig.cm1MUV}
\end{figure}

The effect of introducing the Majorana mass terms can be seen in the
different negative and positive masses as one moves away from
$M_{UV_o}, M_{IR_o}=0$. Due to the different behavior of the
functions $\tilde{S}$ and $\dot{\tilde{S}}$, by inspection, we can
see that the positive and negative roots of Eq.~(\ref{odd_maj}) will
no longer be equal. The two Dirac states have been split into four
Majorana states. These states can still be recognized as two mostly
singlet and two mostly doublet states by comparing their masses to
the charged states. In the following, we will sometimes refer to
these states as singlet or doublet, where it should be understood
that these are not really the original states, but those mixed by
the Higgs. Without the mixing, the coupling between the
singlet-singlet and the doublet-doublet states and the Higgs would
vanish. Therefore, we expect the singlet-singlet coupling to be
suppressed compared with the coupling of the mostly singlet state to
the mostly doublet states. Additionally, looking at
Fig.~\ref{fig.cm1MUV} we see that as expected, the Majorana masses
don't effect the mass of the LOP when the LOP is mainly a doublet
state (purple circles for $c_2>0$), as the convergence of the
different curves corresponding to the different values of $M_{IR_o}$
and $M_{UV_o}$ clearly show.

The behavior of the LOP mass as well as its coupling to the Higgs
may be studied by looking at the roots of Eq.~(\ref{odd_maj}). Using
the small $z$ expansion for $\tilde{S}_{M}$, we obtain
\begin{equation}
z\sim 2~\tilde{k}~ \left(\frac{1}{2}+c_2
\right)~\frac{M_{IR_o}M_{UV_o} -e^{2 c_2 k L}\cos\left[\frac{\lambda
h}{f_h}\right]^2}{e^{2 c_2 k
L}M_{IR_o}+4\frac{(\frac{1}{2}+c_2)}{(\frac{1}{2}-c_2)}M_{UV_o}}~ ,
\label{maj_app}
\end{equation}
which is valid for the case in which at least one of the Majorana
masses is non-vanishing and $c_2 < 0$, in which the singlet becomes
the LOP. This clearly shows that when only one of the Majorana
masses is non-zero, we get a See-Saw effect governing the LOP mass.
This behavior is confirmed in our numerical analysis plotted in
Fig.~\ref{fig.cm1MUV}.

Observe that if only the ultraviolet mass is non-vanishing, the mass
of the LOP (which is mainly a left-handed singlet) is exponentially
suppressed, unless $c_2 \simeq 0$. Indeed, in the limit of vanishing
infrared Majorana masses, Eq.~(\ref{maj_app}) reduces to
\begin{equation}
z\sim ~ -\tilde{k}~ \left(\frac{1}{2}-c_2 \right)~\frac{e^{2 c_2 k
L}\cos\left[\frac{\lambda h}{f_h}\right]^2}{ 2 M_{UV_o}}~.
\label{maj_appmuv}
\end{equation}
As can be seen from Eq.~(\ref{maj_appmuv}), the higher operator
coupling of the LOP to the Higgs is also suppressed for $c_2 < 0$.
We will show that the annihilation cross section for a mainly
singlet state is sufficiently enhanced only when the s-channel Higgs
diagram becomes sizable and therefore, unless $c_2 \simeq 0$, the
Dark Matter density becomes very large compared to the
experimentally observed value.

As both $M_{IR_o}$ and $M_{UV_o}$ are turned on, we see an abrupt
change in the behavior of the mass spectrum for $c_2<0$, which becomes
independent of the exact value of $M_{UV_o}$ and only depends on $\tilde{k}$
and $M_{IR_o}$,
\begin{equation}
z\sim
~\frac{\tilde{k} \; M_{IR_o}}{
2} ~
\left(\frac{1}{2}-c_2 \right) ~.
\label{maj_appmuvmir}
\end{equation}
The LOP mass in this case is of the order of $\tilde{k}$, and does
not show an explicit dependence on the Higgs vaccum expectation
value. Indeed, as can be seen from Eq.~(\ref{maj_app}), the
effective coupling to the Higgs for $c_2 <0$ is exponentially
suppressed. Therefore, as happens in the case of vanishing
$M_{IR_o}$, a good dark matter candidate may only be obtained for
values of $c_2 \gtrsim 0$.

When $c_2\sim 0$, the mass can be approximated by:
\begin{equation}
\label{maj_appc20}
 z\sim
\tilde{k}\frac{M_{IR_o}M_{UV_o}-e^{2 c_2 k L}\cos\left[\frac{\lambda
h}{f_h}\right]^2}{e^{2 c_2 k L}M_{IR_o}+4M_{UV_o}}
\end{equation}
We see that for $M_{IR_o}, ~M_{UV_o}\sim \mathcal{O}(1)$, for values
of $h$ in the linear regime and very small values of $c_2$, we can
get a cancelation resulting in very small LOP masses. This behavior
is clearly portrayed in Fig.~\ref{fig.cm1MUV}.

Finally, let us analyze the case $M_{IR_o} \neq 0$ and $M_{UV_o} =
0$. As $M_{IR_o}$ is turned on, it strongly modifies the spectrum
with respect to the Dirac case for negative values of $c_2$. In this
case, the LOP becomes mostly a right-handed singlet and its mass is
given by
\begin{equation}
z\sim ~ - \tilde{k}~ ( 1 + 2 c_2 )~\frac{ \cos\left[\frac{\lambda
h}{f_h}\right]^2}{ M_{IR_o}}~. \label{maj_appmir}
\end{equation}
As seen in Eq.~(\ref{maj_appmir}), the LOP mass in this case is,
again, of the order of the weak scale but with an explicit Higgs
v.e.v. dependence induced by a higher order operator coupling with a
characteristic scale of the order of the KK masses. Therefore the
coupling of the LOP to the Higgs becomes sizable for KK masses of
the order of the TeV scale, allowing for the possibility of a dark
matter candidate for $c_2 < 0$.

In the following section, we shall perform a more precise
quantitative analysis of the masses and couplings associated
with the annihilation cross section of the singlet state for
both the Majorana and Dirac cases.

\subsection{Couplings of the Odd Leptons}

\subsubsection{Higgs Couplings}

To calculate the couplings of the Majorana and Dirac states with the
Higgs, the profile function of the odd leptons which couple to the
Higgs bosons need to be computed. The mass eigenstate profile
functions are given in terms of combinations of the profile
functions without the Higgs. In the particular case of the neutral
odd leptons these are admixtures of the neutral states belonging to the
bidoublet and the singlet state, with normalization coefficients
$C_2$, $C_3$ and $C_5$.

The boundary conditions determine the coefficients $C_3$ and $C_5$
as functions of $C_2$. The fermion profile functions in the presence
of the Higgs are given by:

\begin{eqnarray}
f_L^2(h)&=&\frac{1}{2} e^{\frac{1}{2} (1 - 2 c_2) k x} \left(e^{2
c_2 k x} \dot{\tilde{S}}_{-M_2}\left(C_2 \left(1 +
\cos\left[\frac{\lambda h}{f_h}\right]\right) -C_3 \left(1 -
\cos\left[\frac{\lambda h}{f_h}\right]\right)\right)\right. \nonumber\\
&&-\left. \sqrt{2} (\tilde{S}_{M_2} - e^{2 c_2 k x} M_{UV_o}
\dot{\tilde{S}}_{-M_2})~ C_5 \sin\left[\frac{\lambda h}{f_h}\right]\right)\label{fl2}\\
&&\nonumber\\
f_L^3(h)&=&\frac{1}{2} e^{\frac{1}{2} (1 - 2 c_2) k x} \left(e^{2
c_2 k x} \dot{\tilde{S}}_{-M_2}\left(C_3 \left(1 +
\cos\left[\frac{\lambda h}{f_h}\right]\right) -
    C_2 \left(1 - \cos\left[\frac{\lambda h}{f_h}\right]\right)\right)\right.\nonumber\\
&&   -\left. \sqrt{2} (\tilde{S}_{M_2} - e^{2 c_2 k x} M_{UV_o}
\dot{\tilde{S}}_{-M_2})~ C_5 \sin\left[\frac{\lambda h}{f_h}\right]\right)\label{fl3}\\
&&\nonumber\\
 f_L^5(h)&=&\frac{1}{2} e^{\frac{1}{2} (1 - 2 c_2) k x}
\left(2 (\tilde{S}_{M_2} - e^{2 c_2 k x} M_{UV_o}
\dot{\tilde{S}}_{-M_2}) ~C_5 \cos\left[\frac{\lambda h}{f_h}\right]
+ \sqrt{2} e^{2 c_2 k x} \dot{\tilde{S}}_{-M_2} (C_2 + C_3)
\sin\left[\frac{\lambda h}{f_h}\right]\right)\nonumber\\
&&\label{fl5}
\end{eqnarray}
\begin{eqnarray}
 f_R^2(h)&=&\frac{1}{2} e^{\frac{1}{2} (1 - 2 c_2) k
x} \left(e^{2 c_2 k x} \tilde{S}_{-M_2}\left(C_2 \left(1 +
\cos\left[\frac{\lambda h}{f_h}\right]\right)-C_3 \left(1 -
\cos\left[\frac{\lambda h}{f_h}\right]\right)\right)\right.
   \nonumber\\
&&   +\left.   \sqrt{2} (e^{2 c_2 k x} M_{UV_o} \tilde{S}_{-M_2} -
\dot{\tilde{S}}_{M_2})~ C_5 \sin\left[\frac{\lambda h}{f_h}\right]\right)\label{fr2}\\
&&\nonumber\\
 f_R^3(h)&=&\frac{1}{2} e^{\frac{1}{2} (1 - 2 c_2) k x}
\left(e^{2 c_2 k x} \tilde{S}_{-M_2}\left( C_3\left(1 +
\cos\left[\frac{\lambda h}{f_h}\right]\right) - C_2 \left(1 -
\cos\left[\frac{\lambda h}{f_h}\right]\right)\right)\right.\nonumber\\
&&   + \left.  \sqrt{2} (e^{2 c_2 k x} M_{UV_o} \tilde{S}_{-M_2} -
\dot{\tilde{S}}_{M_2})~ C_5 \sin\left[\frac{\lambda h}{f_h}\right]\right)\label{fr3}\\
&&\nonumber\\
 f_R^5(h)&=&\frac{1}{2} e^{\frac{1}{2} (1 - 2 c_2) k x}
\left(2 (\dot{\tilde{S}}_{M_2}-e^{2 c_2 k x} M_{UV_o}
\tilde{S}_{-M_2})~ C_5 \cos\left[\frac{\lambda h}{f_h}\right] +
\sqrt{2} e^{2 c_2 k x} \tilde{S}_{-M_2} (C_2 + C_3)
\sin\left[\frac{\lambda h}{f_h}\right]\right)\nonumber\\
&&\label{fr5}
\end{eqnarray}
where the functions $\tilde{S}$ and $\dot{\tilde{S}}$ are functions
of $x_5$ and the masses $z$ of the odd fermions.

\begin{figure}[!t]\centering\scalebox{0.7}[0.7]
{\includegraphics[width=\linewidth,bb=16 16 526 479,
clip=]{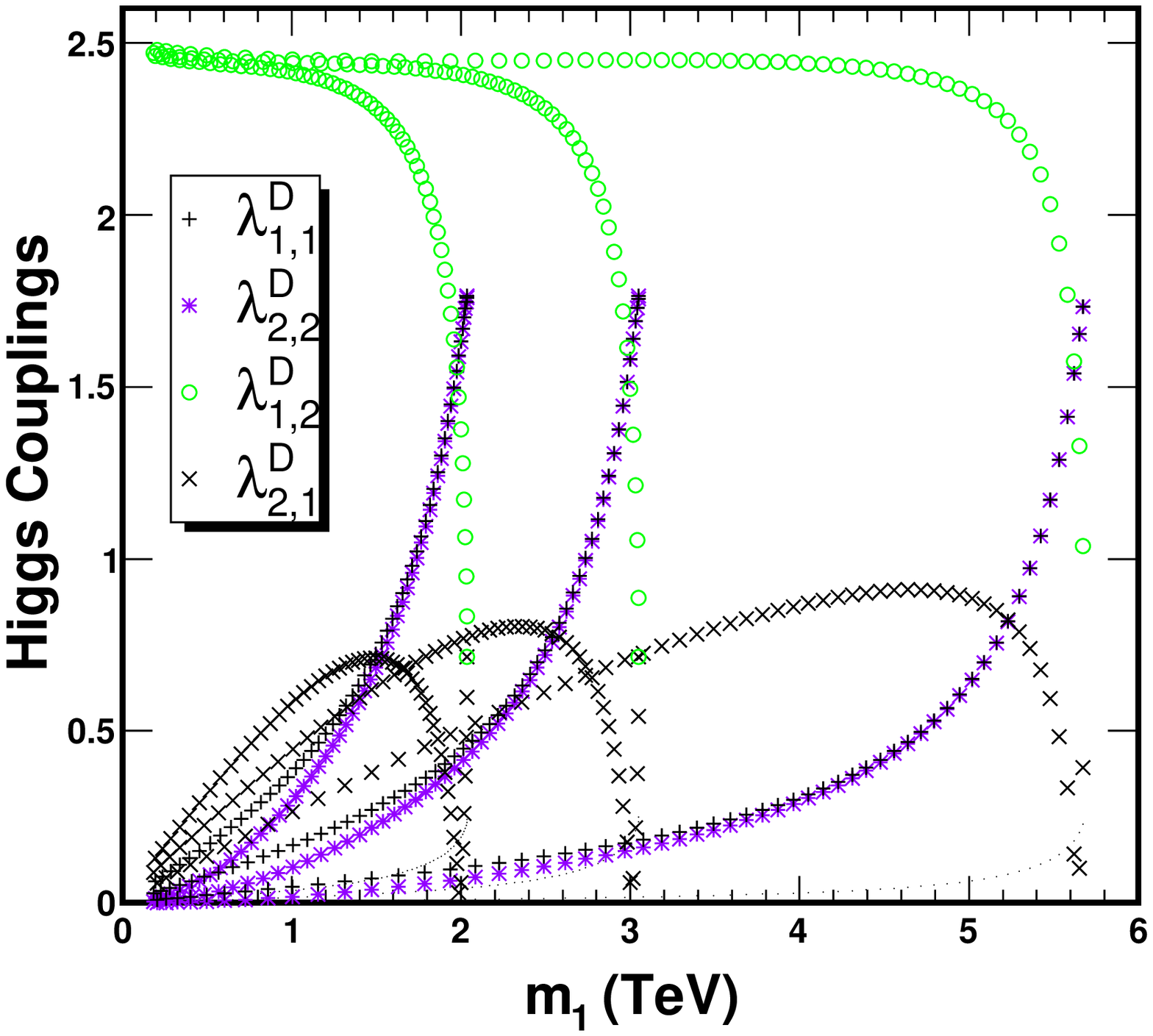}}\caption{\footnotesize{Higgs couplings to the
Dirac particles for three values of $\tilde{k}\sim~1.5,~ 2.2$ and
$3.8$~TeV corresponding to $m_1\sim ~2, ~3$ and $6$~TeV for $c_2=0$
from left to right, as a function of the singlet LOP mass $m_1$. }}
\label{fig.HCD}
\end{figure}

\begin{figure}[!t]\centering\scalebox{0.7}[0.7]
{\includegraphics[width=\linewidth,bb=16 16 526 479,
clip=]{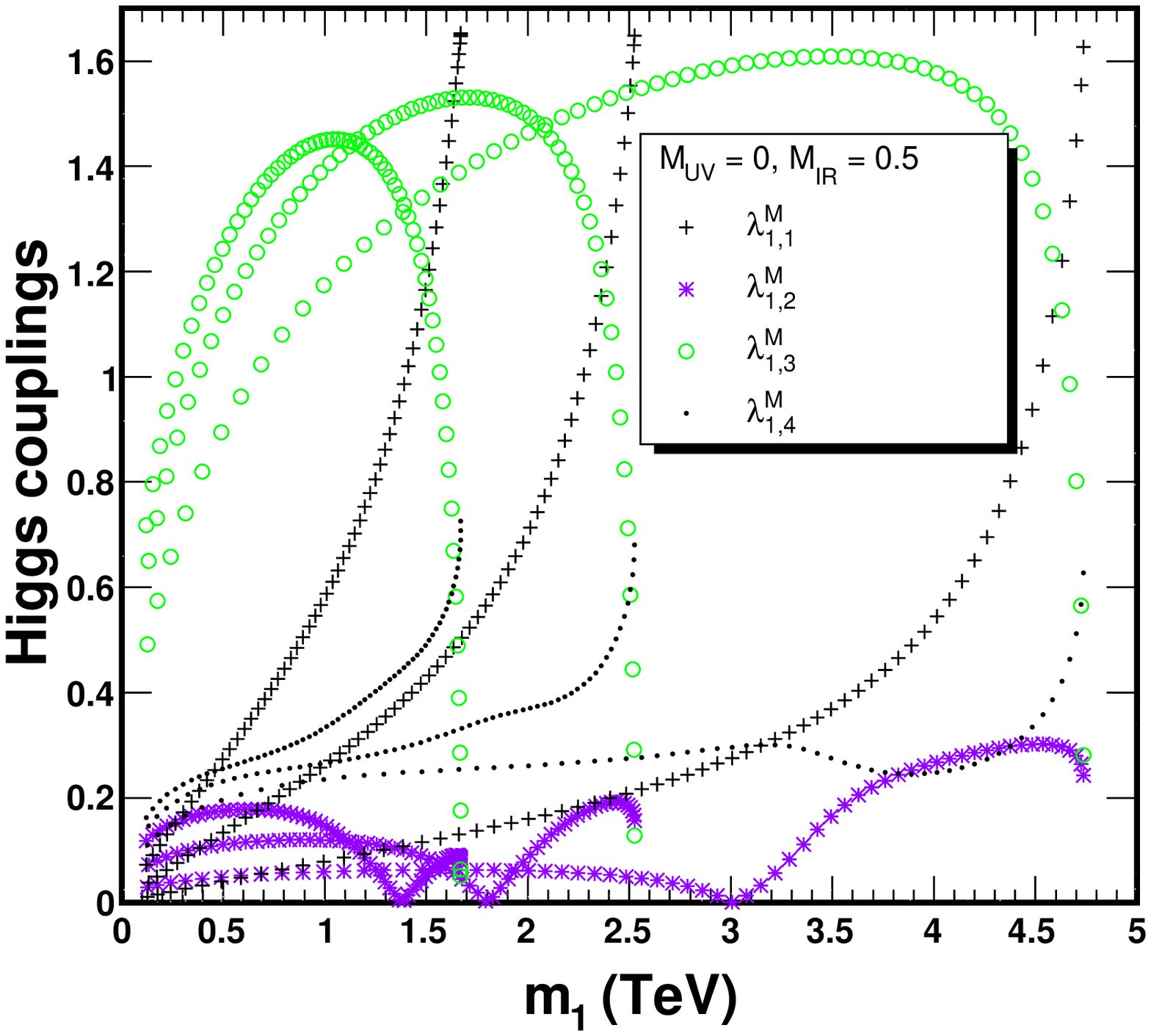}}\caption{\footnotesize{Higgs couplings to
the Majorana particles for three values of $\tilde{k}\sim~1.5,~ 2.2$
and $3.8$~TeV corresponding to $m_1\sim ~1.7,~ 2.5$ and $4.7$~TeV
for  $c_2 = 0$, from left to
right, as a function of  the singlet LOP mass $m_1$. }} \label{fig.HCM0}
\end{figure}

\begin{figure}[!t]\centering\scalebox{0.7}[0.7]
{\includegraphics[width=\linewidth,bb=16 16 526 479,
clip=]{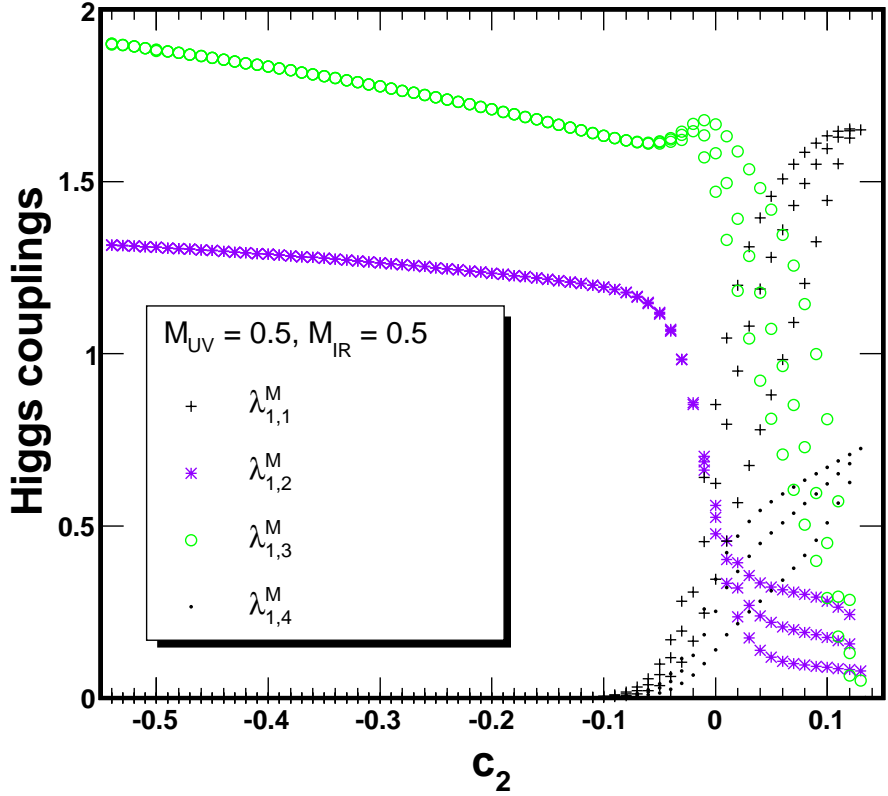}}\caption{\footnotesize{Higgs couplings to
the Majorana particles for three values of $\tilde{k}\sim~1.5,~ 2.2$
and $3.8$~TeV as a function of $c_2$, when the singlet is the LOP.
The smaller values of $\tilde{k}$ correspond to the smaller
couplings (the lower curve).}}\label{fig.HCM5}
\end{figure}

For the doublet and singlet states mixed by the Higgs, a non-trivial
solution may be only obtained if the following relations are
fulfilled.
\begin{eqnarray}
C_3&=&C_2\label{C3}\\
&&\nonumber\\
C_5&=&C_2 \frac{\sqrt{2}e^{2 c_2 k
L}\dot{\tilde{S}}_{-M_2}\cot\left[\frac{\lambda
h}{f_h}\right]_L}{\tilde{S}_{M_2}-e^{2 c_2 k L}
M_{UV_o}\dot{\tilde{S}}_{-M_2}};\label{C5} .\\
&&\nonumber
\end{eqnarray}
For the neutral leptons which decouple from the Higgs, instead, the
following relations are fulfilled:
\begin{eqnarray}
C_3&=&-C_2\label{C32}\\
&&\nonumber\\
C_5&=&0\label{C52}
\end{eqnarray}

This implies that only the symmetric combination of neutral
bidoublet states with coefficients given by Eq.~(\ref{C3}) and
(\ref{C5}), couple to the Higgs. Moreover, the normalization
coefficient $C_2$ may be computed by demanding well normalized
functions, namely,
\begin{equation}\label{C2}
C_2=\left(\int^L_0 \frac{\sum_{i=2,3,5}\left(|f_L^i|^2 +|f_R^i|^2 \right)
}{|C_2|^2}dx\right)^{-1/2}.
\end{equation}
The above definition is appropriate in the Majorana case, in which
the left-handed components of the fermions acquire contributions
from both the original left-handed and the (charge conjugate)
right-handed modes, Eqs.~(\ref{fl2})--(\ref{fr5}). In the Dirac
case, the left-handed and right-handed functions acquire equal
normalizations and therefore the proper factor $C_2$ is equal to the
one computed above divided by $\sqrt{2}$. In the following, we will
keep the above definition of $C_2$ for both the Majorana and Dirac
cases and take care of the proper $\sqrt{2}$ factors explicitly.

To calculate the couplings, we also need the Higgs profile and
normalization:
\begin{eqnarray}
f_h&=&C_h e^{2 k x}\label{higgs}\\
C_h&=&\frac{g_5}{\sqrt{\int_0^L a(x)^{-2} dx}}\label{Ch}
\end{eqnarray}

Defining $\Xi(m_i,m_j)$:
\begin{equation}\label{int}
\Xi(m_i,m_j)=-\frac{e^{-k x} }{2} C_h C_2^*(m_i)
 C_2(m_j) f_h \left[ f_R^{5*}(m_i)\left[f_L^2(m_j) +
      f_L^3(m_j)\right] - \left[f_R^{2*}(m_i)+
      f_R^{3*}(m_i)\right] f_L^5(m_j)\right]
\end{equation}
the left-right couplings of the Higgs with the different states, $\bar{\Psi}^i_L H \Psi^j$,
in the Majorana and Dirac cases can be written as:
\begin{eqnarray}
\lambda^{M}_{i,j}&=&\int^L_0\left(\Xi(m_i,m_j)+
\Xi^*(m_j,m_i)\right)~dx\label{maj_lam}\\
&&\nonumber\\
\lambda^{D}_{i,j}&=&2\int^L_0\Xi(m_i,m_j)~dx\label{dir_lam}
\end{eqnarray}
the factor of $2$ in the Dirac coupling is due to the definition of
the $C_2$ factor discussed above. Observe that, while in the
Majorana case $\lambda^M_{ij} = \lambda^{M\;*}_{ji}$, there is no
such relation in the Dirac case.

The different couplings are plotted in
Figs.~\ref{fig.HCD}~--~\ref{fig.HCM5}. For the Dirac case and $c_2
\simlt 0$, represented in Fig.~\ref{fig.HCD}, small values of the
masses are obtained for smaller values of $c_2$. The left-right
couplings of the singlet and doublet states, $\lambda^D_{1,2}$
acquire large values for negative values of $c_2$. This stems from
the fact that for this case, the left-handed singlet component is
localized towards the IR brane. As $c_2$ goes to 0, the localization
effects become less pronounced and this coupling starts getting
suppressed. The $\lambda^D_{i,i}$ couplings have the opposite
behavior to the cross couplings. For $c_2$ negative, the mass
difference between the singlet and the doublet state is large, while
their mixing is small. Since the self-couplings of the mass
eigenstates are induced by the product of the singlet and doublet
components of these states, they become very suppressed. However, as
$c_2$ goes to zero, the mass eigenvalues become symmetric and
antisymmetric combinations of the singlet and doublet states, and
the self couplings of the mass eigenstates become large, while the
cross couplings tend to zero.

In the Majorana case with $M_{UV_o}=0$, although the quantitative
values are not the same, the $\lambda^M_{1,1}$ and $\lambda^M_{1,4}$
couplings behave similarly to the $\lambda^D_{1,1}$ couplings, since
$m_1$ and $m_4$ are the two mostly singlet states, which are split
due to the non-zero $M_{IR_o}$. When both the Majorana masses are
non-zero, we see an abrupt change in the behavior of the couplings.
As mentioned before, for $c_2 < 0$ the self-coupling of the lightest
state is exponentially suppressed. This behavior is clearly
demonstrated in the Higgs dependent part of the approximation for
the LOP given in Eq.~(\ref{maj_app}). The couplings of the LOP to
the mostly doublet states, however, continue to be large.

\subsubsection{Couplings to the $Z$ and $W^{\pm}$ Bosons}

\begin{figure}[!t]\centering\scalebox{0.7}[0.7]
{\includegraphics[width=\linewidth,bb=10 16 526 479,
clip=]{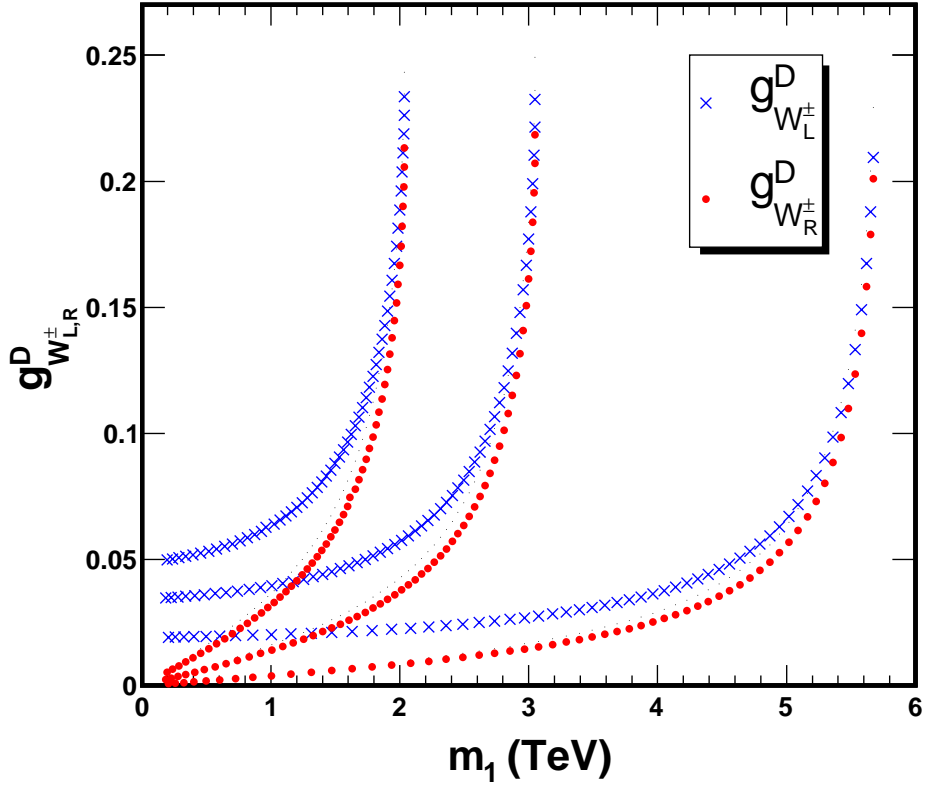}}\caption{\footnotesize{$W^\pm$ couplings to
the Dirac particles for $c_2 < 0$ and three values of
$\tilde{k}\sim~1.5,~ 2.2$ and $3.8$~TeV, which, for $c_2 =0$,
correspond approximately to $m_1\sim~ 2, ~3$ and $5.6$~TeV, from
left to right as a function of the singlet LOP mass. Larger values
of $m_1$ are associated with larger values of $c_2$. }}
\label{fig.GCD}
\end{figure}

\begin{figure}[!t]\centering\scalebox{0.7}[0.7]
{\includegraphics[width=\linewidth,bb=10 16 536 499,
clip=]{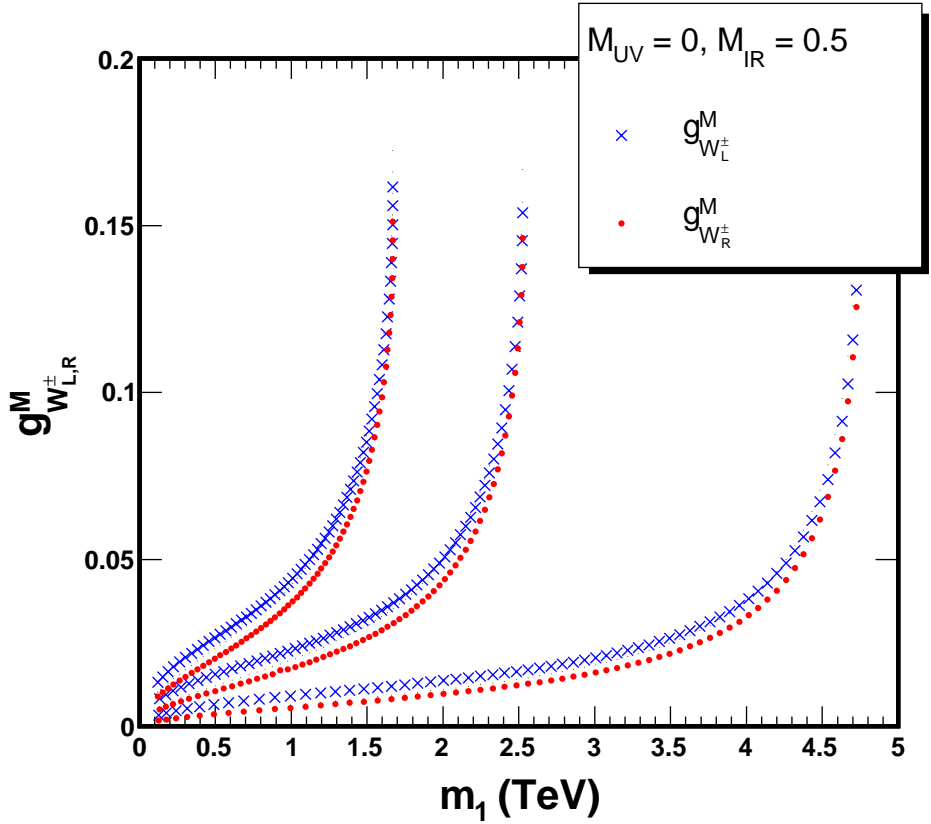}}\caption{\footnotesize{$W^\pm$ couplings of
the Majorana particles for $c_2 < 0$ and three values of
$\tilde{k}\sim~1.5, ~2.2$ and $3.8$~TeV, which, for $c_2 =0$.
correspond to $m_1\sim~ 1.7, ~2.5$ and $4.7$~TeV from left to right
as a function of the singlet LOP mass. Larger values of $m_1$ are
associated with larger values of $c_2$.}} \label{fig.GCM0}
\end{figure}

\begin{figure}[!t]\centering\scalebox{0.7}[0.7]
{\includegraphics[width=\linewidth,bb=10 16 526 479,
clip=]{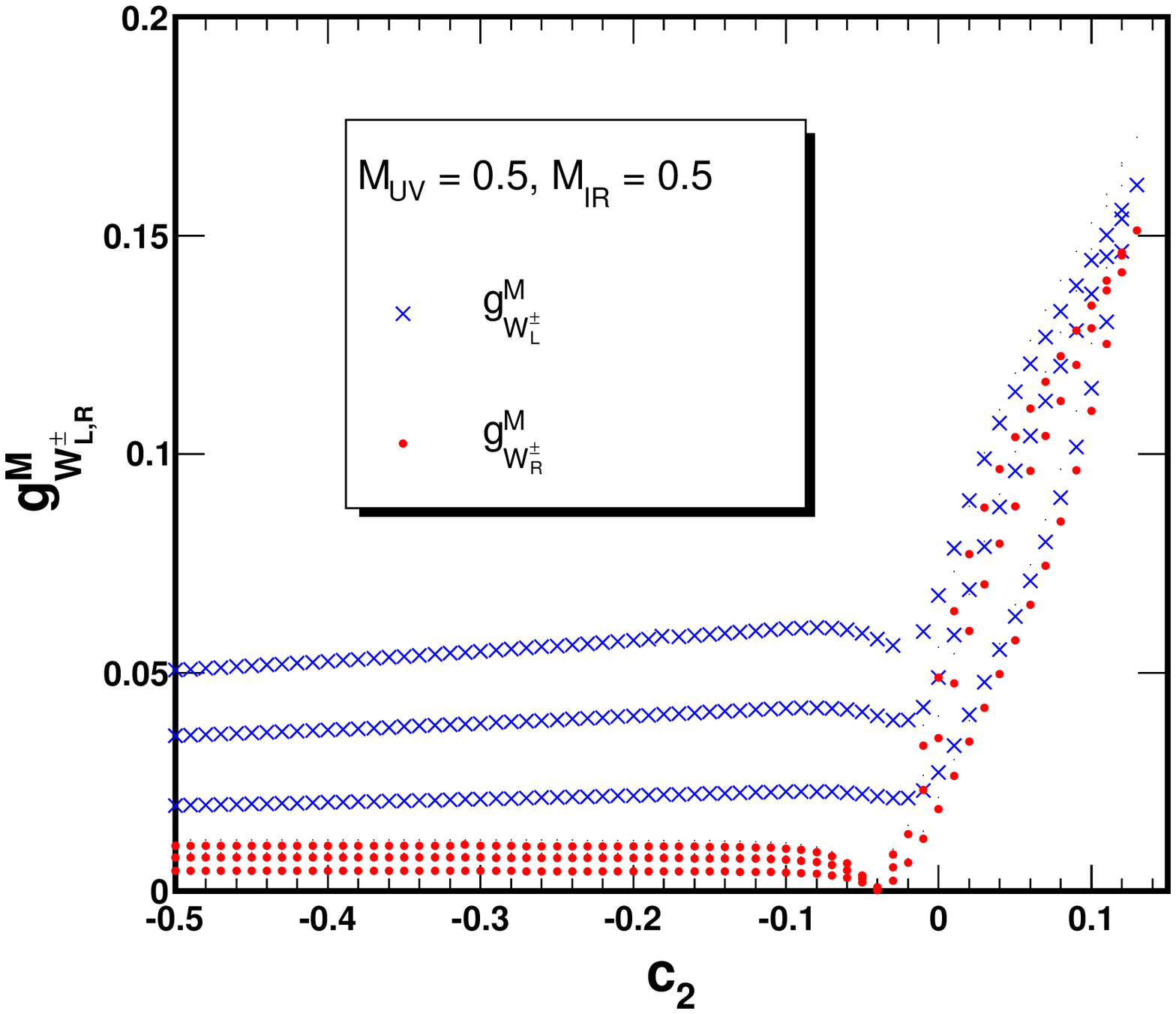}}\caption{\footnotesize{$W^\pm$  couplings to
the Majorana particles for three values of $\tilde{k}\sim~1.5,~ 2.2$
and $3.8$~TeV as a function of $c_2$ for the singlet LOP. The
smaller values of $\tilde{k}$ correspond to the smaller couplings
(the lower curve).}}\label{fig.GCM5}
\end{figure}

The $Z$ couplings to the lepton sector are defined in a similar
manner to the couplings with the quark
sector~\cite{Medina:2007hz},\cite{Carena:2007tn}. However, in the
lepton case $Q_X = 0$ and the neutral states that couple to the
Higgs have $C_3=C_2$, implying that $f^2_{L,R}(h)=f^3_{L,R}(h)$.
Therefore, the $Z$, and all its KK modes, as well as the neutral
components of the $SU(2)_R$ gauge bosons don't have any couplings
with any two of these states. However, the orthogonal neutral state
in the bidoublet, which does not couple to the Higgs has $C_3=-C_2$
and $C_5=0$. Hence, an off-diagonal coupling exists between this
mode, the neutral states that couple to the Higgs and the $Z$.

The $W^\pm$ couple the charged fermions with the neutral components.
In component form, the coupling is between $f^{1,4}_{L,R}(h)$ and
$f^{2,3,5}_{L,R}(h)$. The profile functions and their normalization
coefficients for the neutral components were given in
Eqs.~(\ref{fr5})~--~(\ref{C3}). The charged fermions and the neutral
component of the bidoublet which don't couple to the Higgs state are
both governed by the same five dimensional wave-function, namely:
\begin{eqnarray}
f^{i}_{L}&=&C_{i} e^{\frac{1}{2}(1+2c_2)k
x}\dot{\tilde{S}}_{-M_2}\label{fl4}\\
f^{i}_{R}&=&C_{i} e^{\frac{1}{2}(1+2c_2)k
x}\tilde{S}_{-M_2}\label{fr4}
\end{eqnarray}
These fermion masses are given by the roots of
$\tilde{\dot{S}}_{-M_2}$ and since the Majorana masses don't
influence them, they are always Dirac states. Further, it can be
shown trivially that the $W^-$ coupling to the LOP and the positive
charged state is equal to the coupling of the LOP to $W^+$ and the
negative charged states. In the annihilation cross-section, we will
only be interested in the couplings between the two charged fermions
and the LOP. We will denote these couplings by $g_{W_{L,R}}$. The
expression for these couplings is given in Appendix B. Similarly,
for the $Z$ we will only be interested in the couplings between the
LOP and the $N'$ state, the neutral bidoublet component that does
not mix with the Higgs, and we will denote these couplings by
$g_{Z_{L,R}}$.

The gauge boson couplings are plotted in
Figs.~\ref{fig.GCD}~--~\ref{fig.GCM5}. Again we see that the Dirac
and the Majorana, $M_{UV_o}=0$ couplings behave in a similar way.
The coupling of the mostly singlet state to the charged or neutral
fermion is obtained through the mixing with the bidoublet states. As
discussed before, this mixing is small, for $c_2 <0$, and increases
for larger values of $c_2$. As $c_2$ approaches 0, in the Dirac
case, we expect that since the mixing is maximal, the $W^{\pm}$
couplings should approach the neutrino-lepton SM coupling,
$g_w/\sqrt{2}$, reduced by a factor $1/\sqrt{2}$, due to the mixing
of the singlet with the doublet state, times another factor
$1/\sqrt{2}$ due to the projection of the neutral doublet state on
the $SU(2)$ partner of each of the charged fields. This behavior is
clearly seen in Fig.~\ref{fig.GCD}, where only values of $c_2 < 0$
are plotted and increasing values of $m_1$ are associated with
larger values of $c_2$. For $c_2 <0$, the left-handed states which
are located towards the infrared brane couple more strongly to the
charged states than the right-handed states. Moreover, the couplings
of the $Z$ and the $W^{\pm}$ become proportional to each other, with
a coefficient of proportionality governed by $\cos\theta_W$, namely
\begin{equation}
g_{Z_{L,R}} =
\frac{g_{W_{L,R}}}{\cos\theta_W}
\end{equation}
The additional factor of $\sqrt{2}$ that appears between SM
couplings of neutrinos to the $Z$ and $W$, is not present in this
case.

In the case of zero ultraviolet Majorana mass but non-vanishing
$M_{IR_o}$, depicted in Fig.~\ref{fig.GCM0}, the behavior is similar
to the Dirac case but the couplings are reduced due to the larger
singlet components of the Majorana particles. Also, there is a
sizable reduction of the left-handed couplings due to the larger
right-handed component of the Majorana state.
Observe that as $M_{UV_o}$ is turned on, for $c_2 < 0$, the
couplings to the gauge bosons become independent of $c_2$.

Finally, as $c_2$ becomes positive, the couplings increase, due to
the larger bidoublet component of the LOP.

\subsection{Annihilation Cross Section}

We will denote the neutral states mixed by the Higgs by $N_i$, where
$i=1,2$, or $i=1,2,3,4$ for the two Dirac or four Majorana states
respectively, where $i$ labels the states in increasing order of
their absolute masses. $C^{\pm}$ will denote the charged fermions
and, as said before, $N'$ will denote the bidoublet neutral fermion
which does not couple to the Higgs. The $N_1$ is the LOP, our dark
matter candidate.

\begin{figure}[!t]\centering\scalebox{0.3}[0.3]
{\includegraphics[width=\linewidth,bb=302 82 575
385,clip=]{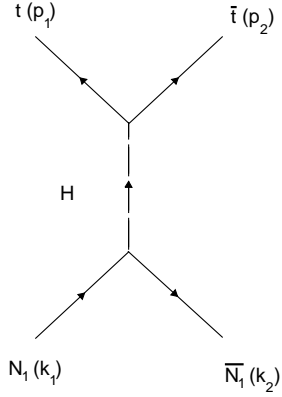}}\caption{\footnotesize{Feynman Diagram for the
process $N_1+\bar{N}_1\rightarrow t+\bar{t}$  }} \label{fig.NNttF}
\end{figure}

\begin{figure}[!t]\centering\scalebox{0.5}[0.5]
{\includegraphics[width=\linewidth,bb=21 52 560 531,
clip=]{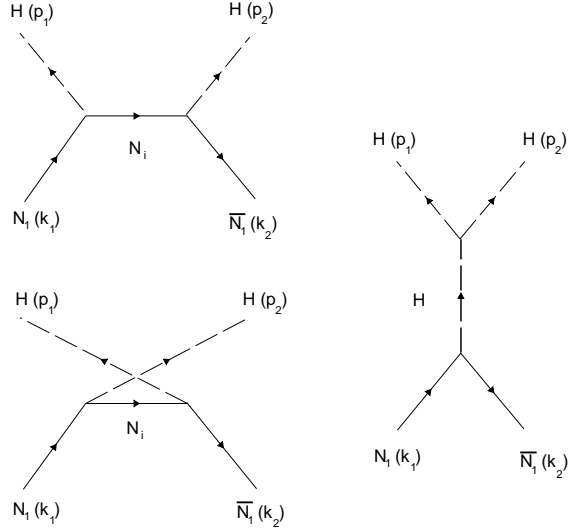}}\caption{\footnotesize{Feynman Diagrams
contributing to the process $N_1+\bar{N}_1\rightarrow H+H$  }}
\label{fig.NNHHF}
\end{figure}

Ignoring co-annihilation effects, we consider the following five
dominant processes for $N_1N_1$ annihilation:
$N_1+\bar{N}_1\rightarrow t~\bar{t}, ~H~H, ~Z~Z$ and $W^{+}~W^{-}$
(observe that due to the cancelation of the Z coupling to the states
that couple to the Higgs, the $Z~H$ annihilation channel is
suppressed). The Feynman diagrams contributing to each of these
processes are shown in Figs.~\ref{fig.NNttF}~--~\ref{fig.NNWWM}. The
virtual $N_i$ exchanges in these diagrams should be understood to be
summed over $i$, where $i$ as noted above runs over the appropriate
index depending on whether we are considering the Dirac or the
Majorana case. The $v$ in the following formulae is the relative
velocity between the initial particles in the center of mass frame.
$\lambda_{Htt}$, $\lambda_{HZZ}$, $\lambda_{HWW}$ and $\lambda_{H}$
are the couplings of the Higgs to the top, the $W^\pm$ and $Z$
bosons, and itself, which were discussed in
Refs.~\cite{Medina:2007hz} and \cite{Carena:2007tn}.

\subsubsection{$N_1+\bar{N}_1\rightarrow
t+\bar{t}$}

Due to the cancelation of the coupling of $N_1$ to the $Z$, the
annihilation into fermion pairs proceeds via an s-channel Higgs
interchange, and is therefore proportional to the corresponding
fermion mass. Therefore, only the top contributes in a relevant way.
The Dirac and Majorana cross-sections are given by the same formula,
but the Higgs coupling should be understood to be the one
appropriate for each case. Assuming $m_1 > m_t$, we obtain:
\begin{equation}\label{NNtt}
<\sigma v>_{tt}=\frac{\lambda_{1,1}^{2}\lambda^2 _{Htt}v^2}{8 \pi
m_1^2}\left(1-\frac{m_t^2}{m_1^2}\right)^{3/2}\left(1-\frac{m_H^2}{4m_1^2}\right)^{-2}
\end{equation}
\subsubsection{$N_1+\bar{N}_1\rightarrow H+H$}

The annihilation into Higgs pairs proceeds via an s-channel Higgs
interchange diagram, which is subdominant, and the t-channel
interchange of the neutral odd fermions. The result, in the limit
$m_H\ll m_1,m_2$, is given by:

\begin{eqnarray}
&&<\sigma v>^D_{HH}=\nonumber\\
&&\nonumber\\
 && \frac{v^2}{8
\pi^2}\left[\frac{\lambda^{D~2}_{1,1}}{16 m_1^2}
\frac{\lambda^2_{H}}{m_1^2}\right.\nonumber\\
&&-\sum_{i}\frac{\lambda^{D}_{1,1}}{4
m_1^2}\frac{\lambda_{H}}{m_i}\left(\left(\lambda^{D~2}_{1,i}+\lambda^{D~2}_{i,1}\right)\frac{m_1}{m_i}\left(1+\frac{m_1^2}{3m_i^2}\right)+2\lambda^D_{1,i}\lambda^D_{i,1}\left(1+\frac{m_1^2}{m_i^2}\right)\right)\nonumber\\
&&+\sum_{i,j}\frac{1}{m_i
m_j}\left(1+\frac{m_1}{m_i}\right)^{-2}\left(1+\frac{m_1}{m_j}\right)^{-2}\left(2\lambda^{D}_{1,j}\lambda^{D}_{j,1}\left(1+\frac{m_1^2}{m_j^2}\right)\left(2\lambda^D_{1,i}\lambda^D_{i,1}\left(1+\frac{m_1^2}{m_i^2}\right)\right.\right.\nonumber\\
&&+\left.\left(\lambda^{D~2}_{1,i}+\lambda^{D~2}_{i,1}\right)\frac{m_1}{m_i}\left(1+\frac{m_1}{3m_i^2}\right)\right)+\left(\lambda^{D~2}_{1,j}+\lambda^{D~2}_{j,1}\right)\frac{m_1}{m_j}\left(\left(\lambda^{D~2}_{1,i}+\lambda^{D~2}_{i,1}\right)\frac{m_1}{m_i}\left(1+\frac{m_1}{3m_i^2}\left(1+\frac{m_i^2}{m_j^2}\right)\right.\right.\nonumber\\
&&+\left.\left.\left.\frac{m_1^4}{m_i^2m_j^2}\right)+2\lambda^D_{1,i}\lambda^D_{i,1}\left(1+\frac{m_1^2}{m_i^2}\right)\left(1+\frac{m_1^2}{3m_j^2}\right)\right)\right]\label{HHDirac}
\end{eqnarray}
where we have taken the couplings to be real. The sum runs over the
two Dirac states labeled by their masses, $m_1$ and $m_2$. In the
case of real couplings, the Majorana cross-section can be simply
seen from the above with the replacement
$\lambda^{M}_{i,j}=1/2(\lambda^D_{i,j}+\lambda^D_{j,i})$, and the
indices now run over the four Majorana states:
\begin{eqnarray}
&&<\sigma v>^M_{HH}=\nonumber\\
&&\nonumber\\
 && \frac{v^2}{2
\pi^2}\left[\frac{\lambda^{M~2}_{1,1}}{64 m_1^2}
\frac{\lambda^2_{H}}{m_1^2}\right.\nonumber\\
&&-\sum_{i}\frac{\lambda^{M}_{1,1}\lambda^{M~2}_{1,i}}{8
m_1^2}\frac{\lambda_{H}}{m_i}\left(\left(1+\frac{m^2_1}{m^2_i}\right)+\frac{m_1}{m_i}\left(1+\frac{m^2_1}{3m_i^2}\right)\right)\nonumber\\
&&+\sum_{i,j}\frac{\lambda^{M~2}_{1,j}\lambda^{M~2}_{1,i}}{m_i
m_j}\left(1+\frac{m_1}{m_i}\right)^{-2}\left(1+\frac{m_1}{m_j}\right)^{-2}\left(\left(1+\frac{m_1^2}{m_j^2}\right)\left(\left(1+\frac{m_1^2}{m_i^2}\right)+\frac{m_1}{m_i}\left(1+\frac{m^2_1}{3m^2_i}\right)\right)\right.\nonumber\\
&&\left.+\frac{m_1}{m_j}\left(\frac{m_1}{m_i}\left(1+\frac{m_1}{3m_i^2}\left(1+\frac{m_i^2}{m_j^2}\right)+\frac{m_1^4}{m_i^2m_j^2}\right)+\left(1+\frac{m_1^2}{m_i^2}\right)\left(1+\frac{m_1^2}{3m_j^2}\right)\right)\right]\label{HHMaj}
\end{eqnarray}

\subsubsection{$N_1+\bar{N}_1\rightarrow
W^{+} W^{-}$, $Z+Z$}

\begin{figure}[!t]\centering\scalebox{0.5}[0.5]
{\includegraphics[width=\linewidth,bb=3 36 354 359,
clip=]{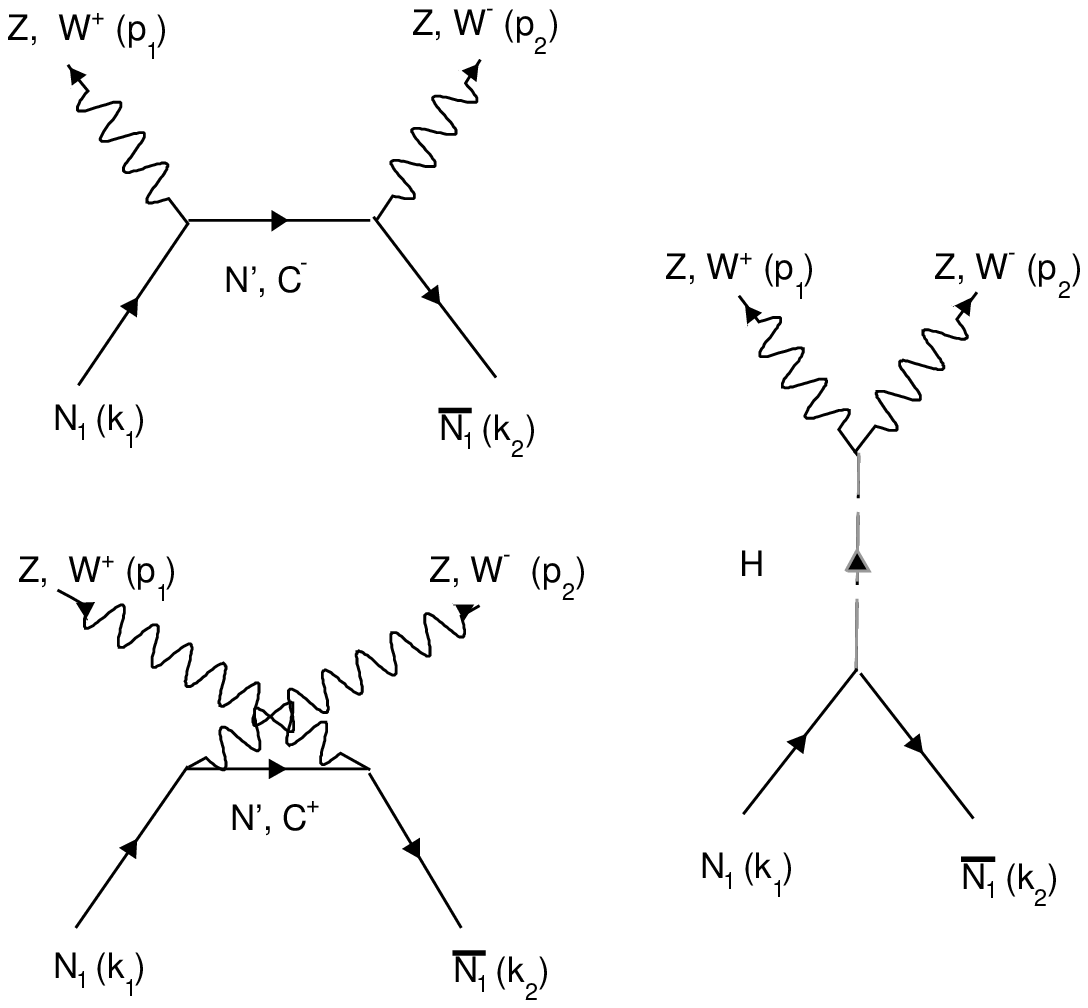}}\caption{\footnotesize{Feynman Diagrams
contributing to the process $N_1+\bar{N}_1\rightarrow W^{+} W^{-},
Z+Z$. The intermediate state is either the charged fermion $C^{\pm}$
for the $W^\pm$ case, or the orthogonal bidoublet, $N'$ for the $Z$.
}} \label{fig.NNWWF}
\end{figure}

\begin{figure}[!t]\centering\scalebox{0.6}[0.7]
{\includegraphics[width=\linewidth,bb=9 191 430 362,
clip=]{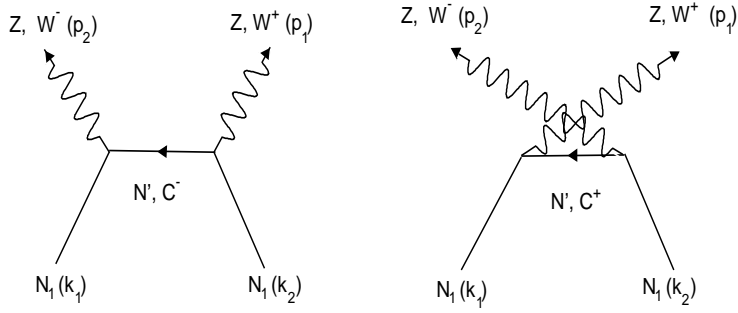}}\caption{\footnotesize{Additional Feynman Diagrams
contributing to the process $N_1+\bar{N}_1\rightarrow W^{+} W^{-},
Z+Z$ for the Majorana case. The intermediate state is either the
charged fermion $C^{\pm}$ for the $W^\pm$ case, or the orthogonal
bidoublet, $N'$ for the $Z$. }} \label{fig.NNWWM}
\end{figure}

The annihilation into the $W^{\pm}$ and $Z$ gauge bosons also
proceeds via the s-channel interchange of a Higgs, plus the
t-channel interchange of the charged fermion $C^{\pm}$ and the
neutral fermion $N'$, respectively. In the formula below, the label
$G$ corresponds to either the $W^\pm$ or $Z$ gauge bosons, and
$\alpha=1$ for the $W^+W^-$ cross-section and $\alpha=1/2$ for the
$ZZ$ case. The diagrams contributing to the process in the Dirac
case are given in Fig.~\ref{fig.NNWWF}. For the Majorana case, two
additional diagrams contribute, and are given in
Fig.~\ref{fig.NNWWM}. Using the properties of the Majorana
couplings, one can demonstrate that these new diagrams are equal to
the ones associated to the cross diagrams in the amplitudes for the
annihilation into the $W^{\pm}$ (due to the interchange of the
fermion of opposite charge) and $Z$ gauge bosons. Therefore, the
cross-section is given by the same formula for both the Dirac and
Majorana cases, but with appropriate couplings, and a factor
$\beta=2$ for the Majorana case, and $\beta=1$ for the Dirac case.
Although in the numerical analysis the full annihilation cross
section was used, for simplicity, we will only quote the
cross-section for the longitudinal modes, in the limit $
m_W,m_Z<m_H<<m_1$:
\begin{eqnarray}
<\sigma v>_{GG}&=& \alpha\left[ \frac{1}{4 \pi
m_G^2}\frac{m_1^4}{m_f^4}\left(g_{G_L}^2-g_{G_R}^2\right)^2\left(1+\frac{m_1^2}{m_f^2}\right)^{-2}\right.\nonumber\\
&&+~\frac{m_1^2 v^2}{\pi m_G^4}\left[\frac{
\lambda^{2}_{1,1}}{16}\frac{\lambda_{HGG}^2}{m_1^2}\right.\nonumber\\
&&-~\beta~\lambda^{}_{1,1}\frac{\lambda_{HGG}}{m_f}\left(1+\frac{m_1^2}{m_f^2}\right)^{-2}\left(g_{G_L}g_{G_R}\left(1+\frac{m_1^2}{3 m_f^2}\right)-\left(g^2_{G_L}+g^2_{G_R}\right)\frac{m_1}{3m_f}\left(1+\frac{m_1^2}{m_f^2}\right)\right)\nonumber\\
&&+~\beta^2~\frac{2}{3}\frac{m_1^2}{m_f^2}\left(1+\frac{m_1^2}{m_f^2}\right)^{-4}\left(6g^2_{G_L}g^2_{G_R}\left(1+\frac{m_1^2}{m_f^2}+\frac{5m_1^4}{3m_f^4}+\frac{1}{3}\frac{m_1^6}{m_f^6}\right)-\frac{m_1}{m_f}\left(1+\frac{m_1^2}{m_f^2}\right)^2\right.\nonumber\\
&&\qquad\left.\left.\left.\times\left(4g_{G_L}g_{G_R}\left(g^2_{G_L}+g^2_{G_R}\right)-\left(g^4_{G_L}+g^4_{G_R}\right)\frac{m_1}{m_f}\right)\right)\right]\right]
\end{eqnarray}

\begin{figure}[!t]\centering\scalebox{0.65}[0.65]
{\includegraphics[width=\linewidth,bb=8 11 527 470,
clip=]{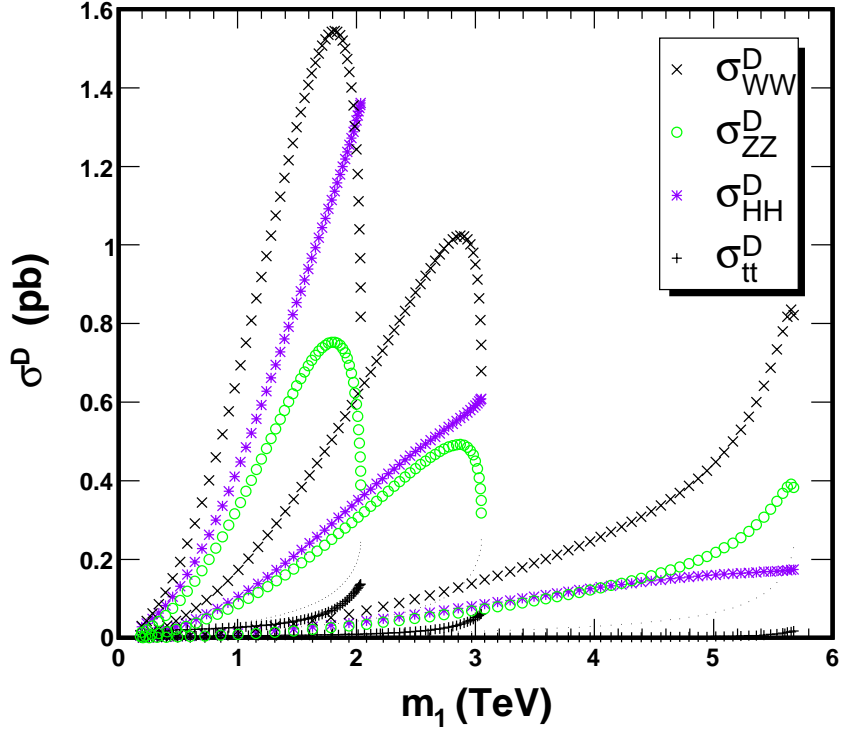}}\caption{\footnotesize{The cross-section
contributions to the annihilation of the Dirac LOP
 for (from top to bottom) $\tilde{k} = 1.5$, 2.2 and 3.8~TeV.}}
\label{fig.CSD}
\end{figure}

\begin{figure}[!t]\centering\scalebox{0.65}[0.65]
{\includegraphics[width=\linewidth,bb=8 11 527 470,
clip=]{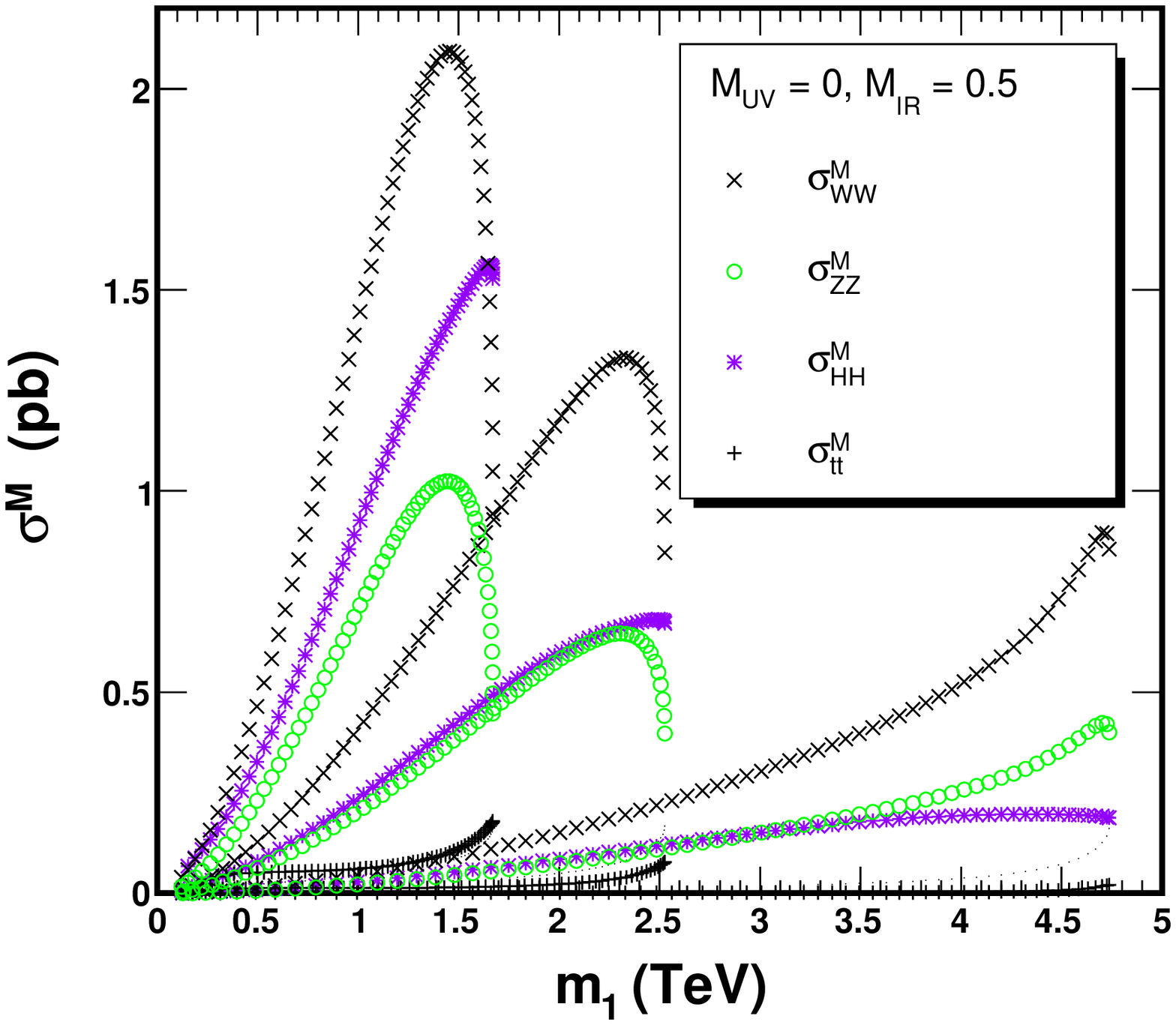}}\caption{\footnotesize{The cross-section
contributions to the annihilation of the Majorana LOP with $M_{IR_o}
= 0.5$ and $M_{UV_o} = 0$,
 for (from top to bottom) $\tilde{k} = 1.5$, 2.2 and 3.8~TeV.}}
\label{fig.CSM0}
\end{figure}

\begin{figure}[!t]\centering\scalebox{0.65}[0.65]
{\includegraphics[width=\linewidth,bb=8 11 527 470,
clip=]{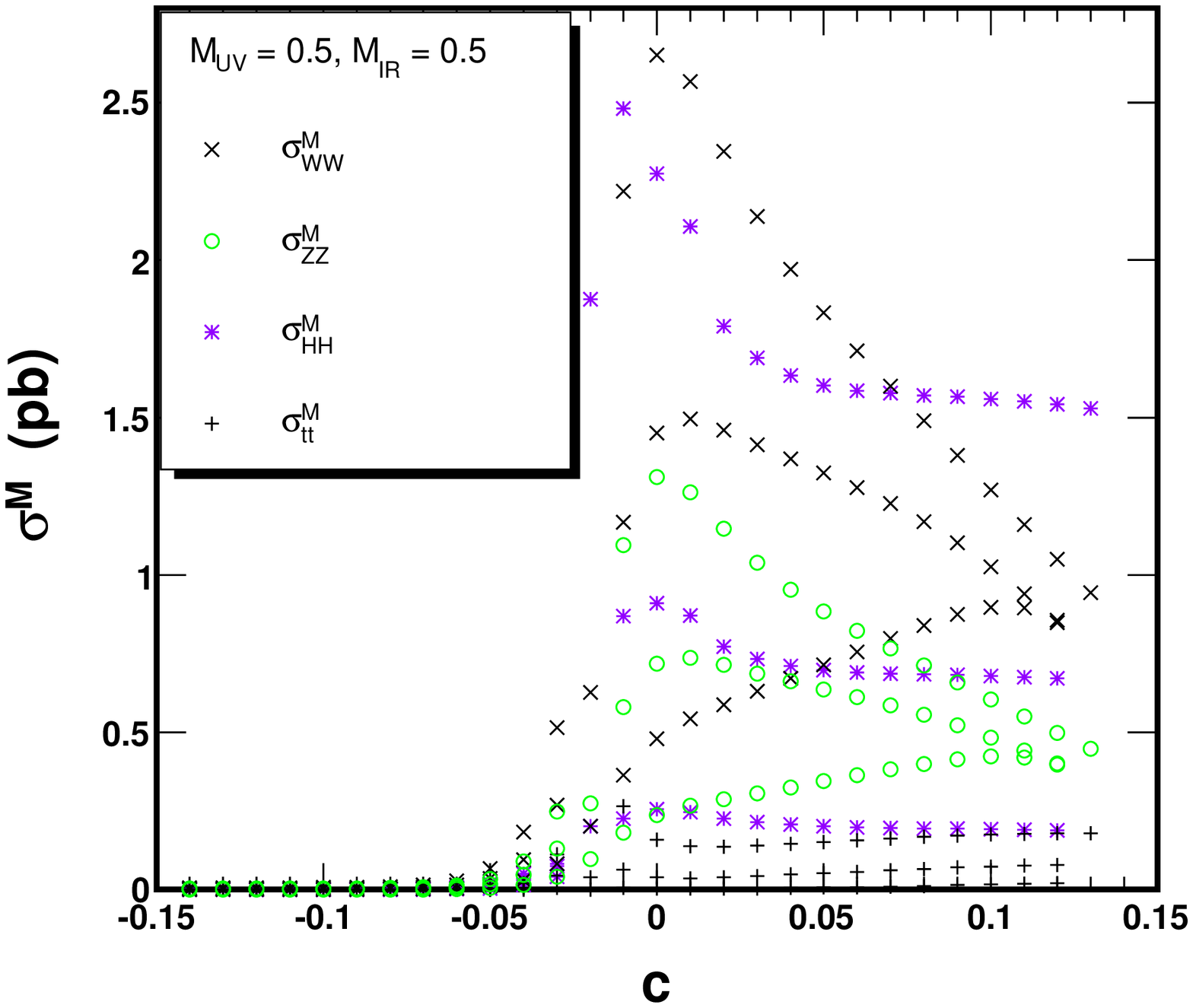}}\caption{\footnotesize{The cross-section
contributions to the annihilation of the Majorana LOP with $M_{IR_o}
= 0.5$ and $M_{UV_o} = 0.5$,
 for (from top to bottom) $\tilde{k} = 1.5$, 2.2 and 3.8~TeV.}}
\label{fig.CSM5}
\end{figure}

These cross-sections are plotted in
Figs.~\ref{fig.CSD}--\ref{fig.CSM5}. For negative $c_2$ (smaller
values of $m_1$), we observe an interesting correlation between the
annihilation cross sections into $W^{\pm}$, $Z$ and Higgs pairs. We
see a dominance of the longitudinal modes for this range of values
of $c_2$ and the magnitudes of the $W^\pm$, $Z~Z$ and $H~H$
cross-sections obey the $2:1:1$ behavior expected due to the
Goldstone equivalence theorem. For larger values of $c_2$, the
bidoublet component of the LOP increases and the transverse
components of the gauge bosons are no longer  subdominant in their
contribution to the annihilation cross section.

Our extensive numerical and analytic study showed that for negative
$c_2$, the major contributions to the $W^\pm$ and $Z~Z$
cross-sections are due to the s-channel Higgs exchange. In the $H~H$
cross-section, this is matched by the contribution from the virtual
exchange of the $N_i$ in the t-channel. Therefore, as emphasized
before, for $c_2 < 0$, a sizable annihilation cross section may only
be obtained when the Higgs coupling to the LOP becomes of order one.

\subsection{Dark Matter Density}

We shall follow the standard formalism for the calculation of the
thermal dark matter
density~\cite{Kolb:1990vq},~\cite{Griest:1990kh}. In calculating the
annihilation cross-sections, we used the non-relativistic
approximation for the initial particles. The cross-section used in
calculating the dark matter density is the sum of all the different
contributions in the previous sections and will be denoted by
$<\sigma v>_T$, and $x_F=m/T_F$ as usual.
The relative velocity is related to the freeze-out temperature:
\begin{eqnarray}
<v^2>_{rel}&=&\frac{6}{x_F}~, \label{vx}\\
&&\nonumber\\
 x_F&=& \log\left(c\left(c+2\right)\sqrt{\frac{90 \pi}{x_F
 g*}}\frac{g_0}{2\pi^3}m_1 M_{Pl}<\sigma
 v>_{T}\right)~.\label{xF}
\end{eqnarray}
The non-relativistic expansion of the thermal annihilation
cross section may be expressed as
\begin{equation}
< \sigma v >_T  = \sigma_0 + \sigma_1 <v^2> \simeq \sigma_0 + 6 \;
\sigma_1/x_F.
\end{equation}
The dark matter density is then given by
\begin{eqnarray}
 \Omega_{DM}&=&\frac{\gamma s_0 x_F}{\rho_c M_{Pl}
(\sigma_0 + 3 \sigma_1/x_F)}
\sqrt{\frac{45}{\pi g*}}\label{Om},
 \end{eqnarray}
where $c=1/2$, $M_{Pl}=1.2\times 10^{19}$ GeV, $g*=112$,
 $s_0=2889.2/\mbox{cm}^3$, $\rho_c=5.3\times
10^{-6}\mbox{GeV}/\mbox{cm}^3$, $g_0=2$ is the degrees of freedom of
our dark matter candidate and $\gamma=2$ or $1$ to account for the
antiparticles for the Dirac and Majorana case respectively. We take
$g_0=2$ in the Dirac case since we compute the density of the
particle and antiparticle separately. The factor $\gamma=2$ in the
relic density then accounts for the duplication of the density from
both the $N_1$ particle and the antiparticle in this case.

\begin{figure}[!t]\centering\scalebox{0.65}[0.65]
{\includegraphics[width=\linewidth,bb=11 21 518 507,
clip=]{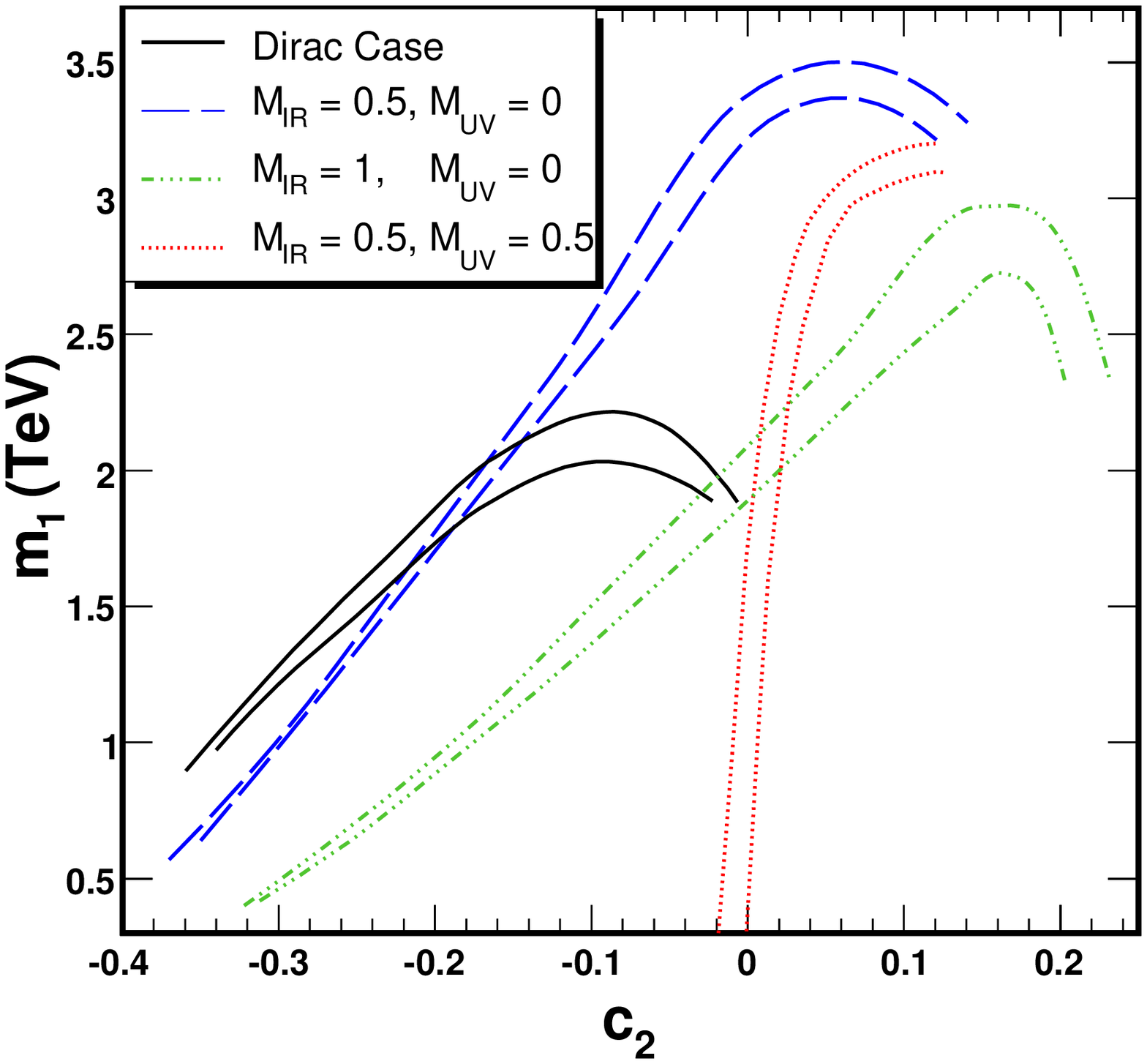}}\caption{\footnotesize{Parametric plot of $m_1$,
the mass of the LOP, versus $c_2$, the localization parameter of the
odd fermions,  when
$\Omega_{DM}\sim0.23\pm 0.1$. The two lines for each value of
the Majorana masses are associated with the upper and lower
bound on $\Omega_{DM}$. }} \label{fig.cm1all}
\end{figure}

To check the veracity of our calculations, we extensively studied
the limit of the Majorana case reducing to the Dirac case as
$M_{IR_o}$ and $M_{UV_o}$ go to 0. As the Majorana masses go to 0,
the two singlet masses start getting degenerate in mass, and the
$N_1~ N_1$ and the $N_2~N_2$ cross-sections become equal. To
properly analyze this limit, then, we must take co-annihilation
between the lightest Majorana sates into
account~\cite{Griest:1990kh}. One can check that the coupling of the
Higgs to each of the degenerate LOP Majorana states, $H N_iN_i$,
becomes equal to the one of the Higgs to the LOP in the Dirac case.
Moreover, the cross coupling of the Higgs, $H N_1 N_2$, vanishes
identically in the limit of vanishing Majorana masses. Further
simple relations exist between the Higgs couplings to fermions in
the Majorana and Dirac cases. One can check that due to these
relations the annihilation cross section into Higgs states in the
Majorana case $N_i N_i \to H~H$ become the same as the $N_1
\bar{N}_1 \to H~H$ cross section in the Dirac case. The same happens
in the case of annihilation into fermions.

In the case of gauge bosons the situation is more complicated. As
noted in calculating the couplings, the gauge boson couplings in the
Majorana case are reduced by a factor $1/\sqrt{2}$. This implies
that for the $W^\pm$ and $Z~Z$, the interference between the t and
s-channel diagrams is reduced by $1/2$ and the t-channel diagrams by
$1/4$ compared to the Dirac cross-section case. These factors are
exactly compensated for by the extra diagrams that contribute in the
Majorana case (Fig.~\ref{fig.NNWWM}), and therefore one obtains that
the equality of annihilation cross sections defined above for the
Higgs final states extends to all final states. We also verified
that the $N_1~N_2$ annihilation cross-section is exactly 0 in this
limit. Including co-annihilation between the two Majorana
states~\cite{Griest:1990kh}, the effective degrees of freedom are
then 4, and the effective cross-section is $\frac{1}{2}~ \sigma^D$,
where $\sigma^D$ is the annihilation cross section between $N_1$ and
its antiparticle in the Dirac case. Therefore, in this limit the
dark matter density due to the Dirac and Majorana cases in the limit
$M_{UV_o,IR_o} \to 0$ is exactly the same, as expected.

We plotted the values of $c_2$ and $m_1$ leading to the correct dark
matter density in both the Dirac and the Majorana cases in
Fig.~\ref{fig.cm1all}. We restricted the values of $\tilde{k} \simgt
1.2$~TeV, for which the SM Higgs couplings are close to their SM
values. This provides a lower bound on the LOP mass and on the value
of $c_2$. The bands in this figure represent the relic density
uncertainty. We see that in the Dirac case, we can only obtain the
correct dark matter density for values of $ 2$ TeV $ \simgt
m_1\simgt 1$~TeV and $c_2\simgt -0.4$, consistent with the exchange
symmetry. The value of $\tilde{k}$ is correlated with the value of
$m_1$, and is constrained to be in the range $2$ TeV $ \simgt
\tilde{k} \simgt 1.2$~TeV for the above range of masses.

In the Majorana case, for $M_{UV_o}=0$, we are able to obtain the
correct dark matter density for a larger range of $m_1$ values,
which extend from about 3~TeV up to the lowest values of about
$m_1\sim 500$~GeV, and for the range of negative values of $c_2
\simgt -0.4$. The values of $m_1$ are, again, correlated with the
values of $\tilde{k}$, which is in the range $3$ TeV $\simgt
\tilde{k} \simgt 1.2$~TeV. For $M_{UV_o}\neq~0$, instead, we see
that we can only obtain the correct dark matter density for
$c_2\simgt 0$. Due to the effect of the Majorana masses, the singlet
LOP state mass is still significantly smaller than the doublet mass
in this case and therefore co-annihilation effects may be ignored.

The values of $c_2$ obtained for the case of vanishing ultraviolet
Majorana masses are fully compatible with the identification of the
LOP with the odd partner of one of the right-handed neutrinos.
Indeed, as can be seen from Fig.~\ref{fig.n1}, for values of $c_1
\simeq 0.5$--0.7, the proper neutrino masses are obtained for values
of $-0.4 \simlt c_2 \simlt -0.1$, for which a proper dark matter
candidate may be obtained, with a mass $0.5$ TeV $\simlt m_1 \simlt
2.5$~TeV.

The situation is different for non-vanishing values of the ultraviolet
Majorana mass $M_{UV_o}$. In this case, a proper dark matter candidate
may only be obtained for values of $c_2 \simgt 0$. Such values of
$c_2$ are incompatible with the exchange symmetry if the $c_1$'s of
the three generations are approximately the same, as assumed in this
article. Therefore, a proper dark matter candidate would demand that
the Majorana mass at the UV brane is either zero or smaller than
$\exp(2 c_2 k \; L)$, since in such a case, according to Eq.~(\ref{maj_app}),
masses of the order of the weak scale and couplings to the Higgs
of order one would be obtained.

Observe that the difficulty in obtaining a proper dark matter
candidate for $M_{UV_o} \neq 0$ stems from the fact that we have
assumed equal bulk mass parameters for the fermions interchanged by
the $Z_2$ exchanged symmetry introduced in
Refs.~\cite{Panico:2008bx}. Although this is an attractive
possibility, since it allows a connection between the dark matter
properties and the neutrino masses, this does not need to be the
case. Values of $c_2 \simgt 0$ would be allowed in the more general
case, or for a more general discrete symmetry. Alternatively, if the
value of $c_1$ of the left-handed leptons was allowed to be
different for the three generations, values of $c_2 \simgt 0$ would
be consistent with values of $c_1$ compatible with the relatively
small electron mass.

\subsection{Direct Dark Matter Detection through Higgs Exchange}

The annihilation cross section for the odd neutrinos becomes of the
proper size for sizable values of the coupling of the odd neutrino
to the Higgs boson, $\lambda_{11} \simeq 0.3$~--~$0.7$, with larger
couplings associated with larger LOP masses. Such large couplings
induce a relatively large scattering cross section of the odd lepton
with nuclei that may be probed at direct dark matter detection
experiments. More quantitatively, the spin-independent elastic
scattering cross-section for an odd lepton scattering off a heavy
nucleus is: \bea \sigma_{SI} = \frac{4m_r^2}{\pi} \left(Z f_p +
(A-Z) f_n\right)^2 \eea where $m_r = \frac{m_N
m_{N_1}}{m_N+m_{N_1}}$ and $m_N$ is the mass of the nucleus. The
factors $f_{p,n}$ are given by \bea f_{p,n} &=& \left(\sum_{q=u,d,s}
f_{T_q}^{(p,n)} \frac{a_q}{m_q} + \frac{2}{27}
f_{TG}^{(p,n)} \sum_{q=c,b,t} \frac{a_q}{m_q} \right) m_{p,n} \\
a_{u,d} &=& - \frac{g_2 m_{u,d} \lambda_{1,1}}{2 m_W m_H^2}
\label{adu:eq}, \eea where the quark form factors are $f_{T_u}^p =
0.020 \pm 0.004, f_{T_d}^p = 0.026 \pm 0.005,  f_{T_s}^p = 0.118 \pm
0.062, f_{TG}^p \approx 0.84,
 f_{T_u}^n = 0.014 \pm 0.003, f_{T_d}^n = 0.036 \pm 0.008,
 f_{T_s}^n = 0.118 \pm 0.062$ and $ f_{TG}^n \approx 0.83 $~\cite{dmdetect}.
Hence, we find that the contribution is

\bea f_{p,n} &\approx& -m_{p,n} \left(
f_{T_u}^{p,n} + f_{T_d}^{p,n} + f_{T_s}^{p,n}
+ \frac{2}{27} f_{TG}^{p,n} \right) \frac{g_2 \lambda_{11}}{2 m_W
m_H^2} \\
&\approx& - 0.2 m_p \frac{g_2 \lambda_{11}}{2 m_W m_H^2} \eea where we
have neglected the differences between the proton and the neutron
mass and have used the fact that the neutron and proton $f_T$
factors are relatively similar. Assuming that the mass of the odd
neutrino is much larger than that of the nucleus we have $m_r \sim
m_N \sim A m_p$ and \bea
\sigma_{SI} &\approx& \frac{4A^2 m_p^2}{\pi} A^2 f_p^2 \\
\Rightarrow \frac{\sigma_{SI}}{A^4} &\approx& \frac{0.04 \;
\lambda_{11}^2  m_p^4 g_2^2}{\pi m_W^2 m_H^4}, \label{sigma_si:eq}
\eea where $\sigma_{SI}/A^4$ is the neutrino-nucleon
spin-independent cross-section. From Eq.~(\ref{sigma_si:eq}), the
spin-independent cross-section scales as $\lambda_{11}^2/m_H^4$ and
therefore direct dark matter detection experiments like CDMS can put
strong constraints on regions of small $m_H$ and large
$\lambda_{11}$.

As discussed above, in the Majorana case, an odd neutrino with a
mass of about 700~GeV and a coupling to the Higgs of about 0.35
leads to an acceptable dark matter density. The spin independent
cross section obtained in such a case for a Higgs mass, $m_H \simeq
130$~GeV, is about $1.4 \times10^{-43}$~cm$^2$. The current limit
coming from CDMS, from a combination of the Ge data and under the
standard assumptions of local dark matter density distribution, is
about 2.5~$\times 10^{-43}$ cm$^2$~\cite{Ahmed:2008eu}. The XENON
experiment puts a slightly weaker limit for this range of LOP
masses~\cite{Angle:2007uj}. Therefore, the predicted spin
independent cross section is only a factor of a few lower than the
current experimental limits.  For larger masses, of about 1~TeV,
the Higgs couplings
grows to about 0.38 and the predicted cross section is therefore
about $1.65 \times 10^{-43}$~cm$^2$, while the CDMS bound is about
two times larger. In the Dirac case, the couplings are about
fifty percent larger than in the Majorana case
and therefore the predicted cross section
for a mass $m_1 \simeq 1$~TeV is slightly above the CDMS
reported bound. One would be able to conclude that the Dirac case
for $m_H \simlt 130$~GeV is
therefore disfavored. There are, however, uncertainties of order of
a few associated with the local dark matter density distribution and
the nuclear form factors which should be taken into account before
ruling out a specific model. It is expected that both the XENON and
CDMS experiments will further improve their sensitivity by about an
order of magnitude by the end of
2009~\cite{xenon100},~\cite{Bruch:2007zz}. Therefore, even
considering possible uncertainties associated with the local density
and the nuclear form factors, the minimal model discussed in this
article should be probed by these experiments in the near future.

Let us comment that the mass of the $N_1$ particle may be in the
appropriate range to provide an explanation of the anomalous excess
in electrons and positrons observed by the
Pamela~\cite{Adriani:2008zr} and ATIC~\cite{:2008zz} experiments.
However, since in these model these particles decay mostly into
Higgs and gauge bosons, if these particles are distributed
throughout the halo of the galaxy, an excess of positrons of the
size observed by these experiments will need a large boost factor
enhancement and would probably lead to an unobserved excess of
antiprotons~\cite{Cirelli:2008pk},\cite{Hall:2008qu}.

\section{Conclusions}

In this article, we have considered the question of incorporating
the charged and neutral leptons in a Gauge-Higgs Unification
scenario based on the gauge group $SO(5) \times U(1)_X$ in warped
extra dimensions. These models are attractive since the $SO(4)
\equiv SU(2)_L \times SU(2)_R$ subgroup of $SO(5)$ incorporates in a
natural way the weak gauge group as well as an appropriate custodial
symmetry group in order to suppress large contributions to the
precision electroweak observables. Moreover, the fifth dimensional
components of $SO(5)/SO(4)$ gauge bosons have the right quantum
numbers to play the role of the Higgs doublet responsible for the
breakdown of the electroweak symmetry.

We have shown that, similar to the quark sector, the leptons can be
incorporated by including the left-handed zero modes in a
fundamental representation of $SO(5)$ and the right-handed charged
leptons in a $\bf{10}$ of $SO(5)$. The model includes right-handed
neutrinos which are singlets under the $SO(4) \times U(1)_X$ group
and can therefore acquire localized Majorana masses on the IR and UV
branes. The simple inclusion of the right-handed neutrinos in the
same multiplet as the charged leptons, fails to produce the correct
lepton masses. The correct charged lepton and neutrino masses may be
obtained from a three multiplet structure similar to the quark case.
The bulk mass parameters $c_1$, $c_2$ and $c_3$ of the left-handed
leptons, right-handed neutrinos and right-handed charged leptons,
respectively acquire values $c_1 \simeq 0.5$~--~0.7, $-0.4 \simlt
c_2 \simlt -0.1$ and $-0.9 < c_3 < -0.5$, where larger negative
values of $c_3$ correspond to the first generation leptons.

We have further investigated the possibility of incorporating a
dark-matter candidate by including an exchange symmetry, under which
all SM leptons multiplets are even, and which ensures the stability
of the lightest odd lepton partner. We therefore analyzed the
possibility of associating the dark matter with the lightest neutral
components of the odd leptons, transforming in the fundamental
representation of $SO(5)$. We have shown that these neutral
components have interesting properties. 
The neutral leptons that couple to the Higgs do not have self
couplings to the $Z$-boson. However, these neutral states couple to
the orthogonal combination of neutral states in the bidoublet and to
the $Z$, as well as to the charged leptons and the $W$-gauge boson.

We computed the couplings of the neutral odd leptons to the gauge
bosons and the Higgs in a functional way and computed the dominant
contributions to the annihilation cross section into Higgs, neutral
and charged gauge bosons and fermions (top-quark) final states. We
considered the cases in which the Majorana masses for the neutral
leptons vanish in both branes (Dirac case) as well as the case in
which at least one of them is non-vanishing. By doing that, we have
shown that in the Dirac case, a proper dark matter candidate may be
obtained for masses of about 1~TeV to 2~TeV and localization
parameter $-0.4 \simlt c_2 \simlt -0.1$ in agreement with the
exchange symmetry. If only the Majorana mass in the infrared brane
is non-vanishing, one obtains lower values of the required
odd-lepton masses, which may be of about $500$~GeV to $2.5$~TeV, and
a range for the localization parameter $-0.4\simlt c_2 \simlt -0.1$.
Finally, in the case that the Ultraviolet Majorana mass is different
from zero, the self-couplings of the LOP with the Higgs is
exponentially suppressed for $c_2 < 0$ and becomes non-vanishing
only for $c_2 \simeq 0$, for which a proper dark matter may be
obtained with a mass similar to the one obtained when only $M_{IR}$
is different from zero. This last possibility is incompatible with
the exchange symmetry and the proper neutrino masses if a common
values of $c_1$ is assumed for the three families.

The collider signatures of these models have been previously
discussed in the literature. The odd leptons introduced in this
article will be hard to detect at collider experiments, since the
masses of the charged and neutral non-LOP odd leptons are above a
few TeV, and they have relatively weak interactions. In all cases, a
proper dark matter candidate is obtained for values of the
self-coupling of $N_1$ to the Higgs of about $0.3$~--~$0.7$, with
larger Higgs couplings corresponding to larger LOP masses, for which
the cross section of the dark matter with nuclei becomes sizable. We
have computed the dark matter cross section with nuclei and have
shown that these models will be probed by the CDMS and XENON direct
dark matter detection experiments in the near future.

\subsection*{\sc Acknowledgments}

We would like to thank Shri Gopalakrishna, Howard Haber, Eduardo
Ponton, Jose Santiago  and Tim Tait for useful discussions and
comments. Work at ANL is supported in part by the US DOE, Div.\ of
HEP, Contract DE-AC02-06CH11357. Fermilab is operated by Fermi
Research Alliance, LLC under Contract No. DE-AC02-07CH11359 with the
United States Department of Energy. We would like to thank the Aspen
Center for Physics and the KITPC, China, where part of this work has
been done.

\newpage
\appendix
{\Large{{\bf APPENDIX}}}

\section{Profile Functions at $h = 0$.}

\renewcommand{\theequation}{A.\arabic{equation}}
\setcounter{equation}{0}  

In the $h=0$ gauge, we redefine $\hat{\psi}=a^{2}(x_{5})\psi$ and we
write our vector-fermionic fields in terms of chiral fields. We can
KK decompose the fermionic chiral components as,
\begin{equation}
\hat{\psi}_{L,R}=\sum_{n=0}^{\infty}\psi^{n}_{L,R}(x^{\mu})\hat{f}_{L,R,n}(x_{5})
\end{equation}
where $\hat{f}$ is normalized by:
\begin{equation} \label{ferm.norm} \int^L_0 dx_5
a^{-1}(x_5)\hat{f}_n\hat{f}_m=\delta_{m,n}.
\end{equation}
Therefore the profile function for the zero mode fermion corresponds
to $a^{-1/2}(x_5)\hat{f}_0$.

From the 5D action, concentrating on the free fermionic fields, we
can derive the following first order coupled equations of motion for
$\hat{f}_{L,R,n}$,
\begin{equation}
\label{1order}
(\partial_{5}+M)\hat{f}_{R,n}=(z/a(x_{5}))\hat{f}_{L,n}; \quad
(\partial_{5}-M)\hat{f}_{L,n}=-(z/a(x_{5}))\hat{f}_{R,n}
\end{equation}

We see from Eq.~(\ref{1order}) that we can redefine
$\tilde{f}_{R,L,n}=e^{- Mx_{5}}\hat{f}_{R,L,n}$ and relate the
opposite chiral component of the same vector-like field by
$\tilde{f}_{R,n}=(-a(x_{5})/z)\partial_{5}\tilde{f}_{L,n}$. For the
left handed field having Dirichlet boundary conditions on the UV
brane, we can derive a second order equation for the chiral
component $\tilde{f}_{L,n}$:
\begin{equation}
\label{feom} \left [\partial_5^2  + \left (\frac{a'}{ a} + 2M
\right)
\partial_5   + \frac{z^2}{ a^2} \right ] \tilde{f}_{L,n} = 0
\end{equation}
the solution of which we shall call $\tilde{S}_{M}(x_5,z)$, with
boundary conditions $\tilde{S}_{M}(0,z) = 0$, $\tilde{S}_{
M}'(0,z) = z$.

Similarly, if the right-handed field fulfills Dirichlet boundary
conditions on the UV-brane, we can redefine $\tilde{f}_{R,L,n}=e^{
Mx_{5}}\hat{f}_{R,L,n}$ and then relate the opposite chirality via
$\tilde{f}_{L,n}=(a(x_{5})/z)\partial_{5}\tilde{f}_{R,n}$. We can
further write the equation of motion for $\tilde{f}_{R,n}$:
\begin{equation}
\label{feom1} \left [\partial_5^2  + \left (\frac{a'}{ a} - 2M
\right)
\partial_5   + \frac{z^2}{ a^2} \right ] \tilde{f}_{R,n} = 0.
\end{equation}
We shall correspondingly denote the solution to this equation with
$\tilde{S}_{-M}(x_5,z)$, fulfilling the boundary conditions
$\tilde{S}_{-M}(0,z) = 0$, $\tilde{S}_{- M}'(0,z) = z$.

The solution to Eq.~(\ref{feom}) is given by~\cite{Pomarol:1999ad}:
\begin{equation}
\label{SM} \tilde{S}_M(x_5,z) = \frac{\pi z}{ 2 k} a^{-c-\frac{1}{
2} }(x_5) \left [
 J_{\frac{1}{2}+c} \left (\frac{z}{ k } \right )      Y_{\frac{1}{ 2}+c} \left ( \frac{z}{ k a(x_5)} \right )
- Y_{\frac{1}{2}+c } \left ( \frac{z}{ k } \right )
J_{\frac{1}{2}+c}  \left ( \frac{z}{ k a(x_5)} \right ) \right ].
\end{equation}
The solution for Eq.~(\ref{feom1}), $\tilde{S}_{-M}$, is given by Eq.~(\ref{SM}) with
the replacement $c\rightarrow -c$.

\section{Coupling of the charged gauge bosons}

\renewcommand{\theequation}{B.\arabic{equation}}
\setcounter{equation}{0}  

Following the notation of Ref.~\cite{Carena:2007tn},
the $W^\pm$ boson profile functions are given by:
\begin{eqnarray}
f^{\hat{1}}_G(h)&=& S(x_5)\left(\cos\left[\frac{\lambda
h}{f_h}\right] C^G_{\hat{1}}
+\frac{1}{\sqrt{2}}\sin\left[\frac{\lambda h}{f_h}\right]
C^G_{1_R}\right)-\frac{1}{\sqrt{2}} C(x_5) \sin\left[\frac{\lambda h}{f_h}\right]C^G_{1_L}\label{f1}\\
&&\nonumber\\
&&\nonumber\\
f^{1_L}_G(h)&=& \frac{1}{2}
\left(S(x_5)\left(\left(\cos\left[\frac{\lambda
h}{f_h}\right]-1\right)C^G_{1_R}+\sqrt{2}\sin\left[\frac{\lambda
h}{f_h}\right] C^G_{\hat{1}} \right)+C(x_5)\left(1+
\cos\left[\frac{\lambda h}{f_h}\right]\right)C^G_{1_L}\right)\nonumber\\
&&\nonumber\\
&&\label{f5}\\
&&\nonumber\\
f^{1_R}_G(h)&=& \frac{1}{2}
\left(S(x_5)\left(\left(\cos\left[\frac{\lambda
h}{f_h}\right]+1\right)C^G_{1_R}-\sqrt{2}\sin\left[\frac{\lambda
h}{f_h}\right] C^G_{\hat{1}} \right)+C(x_5)\left(1-
\cos\left[\frac{\lambda h}{f_h}\right]\right)C^G_{1_L}\right)\nonumber\\
&&\nonumber\\
&&\label{f8}
\end{eqnarray}

The normalization coefficients $C^G_{\hat{1},1_R}$, in terms of
$C^G_{1_L}$ are given by:
\begin{eqnarray}
C^G_{\hat{1}}&=&C^G_{1_L}\frac{-4\cos\left[\frac{\lambda
h}{f_h}\right]_L C(L)^\prime S(L)^\prime+C_h h e^{2 k
L}\sin\left[\frac{\lambda h}{f_h}\right]_L\left(C(L)^\prime
S(L)+C(L)S(L)^\prime\right)}{\sqrt{2}S(L)^\prime\left(C_h h e^{2 k
L}\cos\left[\frac{\lambda h}{f_h}\right]_L
S(L)+2\sin\left[\frac{\lambda h}{f_h}\right]_L S(L)^\prime\right)}\label{C1g}\\
C^G_{1_R}&=& -C^G_{1_L}\frac{C(L)^\prime}{S(L)^\prime}\label{C8g}
\end{eqnarray}

The five dimensional weak coupling is defined as $g_{5w}=g_w
\sqrt{L}$. In terms of these, the Dirac couplings for $W^+$ and $Z$,
(denoted by $G$), are given by:
\begin{eqnarray}
g^D_{G1_{L,R}2_{L,R}}&=&-g_{5w}\int^L_0\left(\vec{f}^{+,0'}_{1_{L,R}}.\left(f^{\hat{1}}_G(h)T^{\hat{1}}+f^{1_L}_G(h)T^{1_L}+f^{1_R}_G(h)T^{1R}\right).\vec{f}^{~0}_{2_{L,R}}\right)dx_5\nonumber\\
&=&-g_{5w}\int^L_0\left(f^{1*}_{L,R}(h)\left(f^5_{L,R}(h)f^{\hat{1}}_G(h)+f^2_{L,R}(h)f^{1_{L}}_G(h)-f^3_{L,R}(h)f^{1_{R}}_G(h)\right)\right)dx_5\nonumber\\
&&\label{WLOP}
\end{eqnarray}
Again, due to our choice of normalization for the coefficient $C_2$,
the factor $\sqrt{2}$, coming from the definition of the $W^{\pm}$
fields, is not present in this expression. The Majorana couplings
are given in terms of the Dirac couplings,
\begin{equation}
g^M=\frac{1}{\sqrt{2}}g^D\label{DtoM},
\end{equation}
with $g^D$ given in Eq.~(\ref{WLOP}).

\end{document}